\documentclass[journal]{IEEEtran}
\IEEEoverridecommandlockouts
\usepackage{amsmath,amsfonts}
\usepackage{algorithm}
\usepackage{algpseudocode}
\usepackage{array}
\usepackage{amssymb}
\usepackage{textcomp}
\usepackage{stfloats}
\usepackage{url}
\usepackage{cite}
\usepackage{verbatim}
\usepackage{tabularx}
\usepackage{booktabs}
\usepackage{graphicx}
\usepackage[
colorlinks=true,
citecolor=green,
linkcolor=blue,  
urlcolor=blue
]{hyperref}
\usepackage{balance}
\newcommand{\tr}{\operatorname{tr}}
\begin{document}
\title{Robust Design of Sensing-Assisted Secure Transmission for RIS-Aided ISAC System under Eavesdropper 3D Positioning Uncertainty}
\author{
	Jie Li,
	Zhendong Li,
	Wen Chen,
	Wei Xu,
	and Ning Wang
	\thanks{Jie Li and Ning Wang are with the School of Electrical and Information Engineering, Zhengzhou University, Zhengzhou 450001, China (e-mail: 15534188602@163.com; ienwang@zzu.edu.cn). 
    Zhendong Li is with the School of Information and Communication Engineering, Xi'an Jiaotong University, Xi'an 710049, China (email: lizhendong@xjtu.edu.cn). 
    Wen Chen is with the School of Integrated Circuits, Shanghai Jiao Tong University, Shanghai 200240, China (e-mail: wenchen@sjtu.edu.cn).
	Wei Xu is with the National Mobile Communications Research Laboratory, Southeast University, Nanjing 211189, China (e-mail: wxu@seu.edu.cn).
}
}
\maketitle

\begin{abstract}
Localization errors in estimating the position of a potential eavesdropper (Eve) pose a critical challenge to secure transmission in reconfigurable intelligent surface (RIS)-assisted integrated sensing and communication (ISAC) systems. To address this issue, this paper proposes a sensing-assisted robust secure transmission framework that exploits ISAC sensing to characterize the positioning uncertainty induced by these errors. The RIS is partitioned into a reflection region and a sensing reception region, supporting sensing and communication through spatial separation. By jointly exploiting observations collected at the base station (BS) and the RIS sensing reception region, maximum likelihood estimation (MLE) is employed to estimate the arrival angles, from which Eve's three-dimensional (3D) position is obtained through cross localization. The resulting localization Cramér–Rao lower bound (CRLB) is then used to construct an ellipsoidal uncertainty region around the estimated position. A two-stage optimization strategy is developed. In Stage I, the sensing beamformer and RIS reflection coefficients are jointly optimized to improve localization accuracy. In Stage II, the resulting positioning uncertainty is incorporated into a robust secure transmission design, where the communication beamformer and RIS reflection coefficients are jointly optimized to maximize the worst-case secrecy rate. Simulation results demonstrate that the proposed approach outperforms benchmark schemes without RIS assistance and with random RIS configurations in terms of secrecy performance. The proposed framework effectively mitigates secrecy performance degradation caused by localization errors in Eve's position.

\end{abstract}


\section{Introduction}
With the continuous advancement of the sixth-generation (6G) wireless networks, future wireless systems are expected to evolve from conventional information transmission toward the integration of communication, sensing, computing, and intelligence~\cite{saad2020}. Integrated sensing and communication (ISAC) enables the sharing of wireless spectrum, radio-frequency (RF) hardware, and signal processing resources while simultaneously supporting high-rate data transmission and target detection, localization, and tracking~\cite{tan2021}. By sharing infrastructure, waveforms, and spectral resources, ISAC can improve the utilization efficiency of communication and sensing resources and reduce the hardware and spectrum overhead associated with conventionally separated systems~\cite{Liu2020,zhang2022}.
Therefore, ISAC has emerged as a fundamental framework for integrating communication and sensing capabilities in future wireless networks.
 
However, despite improving resource utilization, ISAC also introduces new security challenges. Since communication and sensing share wireless resources and signal waveforms, sensitive information may be exposed to interception by non-cooperative nodes. Meanwhile, the sensing capability of ISAC provides an opportunity to detect and localize potential eavesdroppers (Eves), linking sensing performance more closely to physical-layer security. In ISAC-enabled secure communication, Eve is often treated as a non-cooperative node whose exact location is difficult to obtain~\cite{Li2026Joint}. Moreover, practical sensing and channel estimation errors may cause beam misalignment and degrade secrecy performance~\cite{Matsumine2025Physical}. In particular, limited observations and measurement noise introduce uncertainty into Eve's estimated position~\cite{Hua2026Joint}, making it necessary to develop robust secure transmission schemes that explicitly account for localization errors.

Reconfigurable intelligent surfaces (RISs) provide additional spatial degrees of freedom for addressing these challenges. 
An RIS consists of a large number of low-cost and low-power programmable electromagnetic elements whose reflection coefficients can reshape the propagation direction and spatial distribution of wireless signals, thereby improving desired-signal quality, suppressing interference, and enhancing spatial coverage~\cite{huang2020,basar2020,bjornson2022}. 
RISs have been investigated in over-the-air federated learning, where RIS phase shifts are jointly optimized with receiver beamforming to facilitate the aggregation of distributed model updates over wireless channels~\cite{Xu2023EdgeLearning}.
For physical-layer security, RIS-assisted systems can jointly optimize BS active beamforming and RIS passive reflection coefficients to enhance the legitimate-user link while suppressing information leakage toward Eve~\cite{Li2026RIS}. 
The studies in~\cite{chen2020,cui2019} demonstrated that RISs could improve secure transmission performance, while~\cite{wu2021,feng2022} jointly optimized BS beamforming and RIS reflection coefficients to maximize the secrecy rate. 
In~\cite{Shen2023}, the additional spatial degrees of freedom provided by RISs were exploited through joint transmit beamforming and phase-shift optimization to mitigate eavesdropping threats under imperfect CSI of both legitimate and eavesdropping channels. 
Beyond communication enhancement, RISs can also reshape sensing propagation paths to strengthen echoes from non-cooperative targets and improve parameter estimation~\cite{Liu2024SNRCRB}. 
These capabilities make RISs a promising basis for integrating sensing and secure communication in ISAC systems.

In RIS-assisted ISAC systems, RISs have been investigated for sensing illumination, target echo reception, and communication reflection, enabling more integrated sensing and communication. The work in~\cite{liu2023ris} investigated the cooperative enhancement of RIS for sensing and communication, where the RIS reflected BS signals toward non-cooperative targets to enhance echoes and reconfigured its reflection phases to improve legitimate-user reception during communication. Furthermore,~\cite{Wu2021Tutorial,Zhang2024} further explored the application of simultaneously transmitting and reflecting (STAR)-RIS to joint sensing and communication design. By simultaneously controlling transmitted and reflected signals with flexible mode/time configurations, STAR-RISs provided additional spatial degrees of freedom, thereby enhancing the synergy between sensing and communication.
However, existing RIS-assisted secure communication studies either assumed perfect Eve CSI~\cite{Jain2025} or modeled Eve's spatial uncertainty over a predefined region~\cite{Salem2026}. These approaches did not explicitly characterize the location uncertainty arising from practical sensing errors, such as limited observations and estimation noise. To address sensing-related uncertainties, recent studies investigated sensing-assisted secure transmission~\cite{Su2024,Li2026Joint} and robust transmission under eavesdropper location uncertainty~\cite{Hong2026}. In addition, robust secure transmission with imperfect sensing estimates~\cite{Jiang2025RISAssistedIS} and joint time and beamforming optimization for RIS-assisted secure ISAC~\cite{Zhang2026JointTime} have also been studied. However, these two works adopted two-dimensional spatial models and therefore did not explicitly characterize Eve's full three-dimensional (3D) location uncertainty. More importantly, the explicit connection between sensing-derived localization accuracy and a CRLB-based 3D location uncertainty set, and its subsequent incorporation into robust RIS-assisted secure transmission design, remains insufficiently explored. These observations motivate a sensing-driven framework that translates localization accuracy into a 3D uncertainty set for Eve and incorporates this uncertainty into robust secure transmission optimization.

Motivated by the above discussion, this paper proposes a hybrid RIS-assisted secure ISAC system. The RIS panel is partitioned into a reflection region and a sensing reception region, and a two-stage time-division duplex (TDD) frame structure is adopted. During the sensing stage, the BS and the sensing reception region jointly collect the echoes from Eve and estimate its 3D position via bistatic localization. Based on the estimated position and its Cramér-Rao lower bound (CRLB), a location uncertainty ellipsoid is further constructed. During the communication stage, the BS beamformer and RIS reflection coefficients are jointly optimized based on the sensing-assisted location uncertainty set to maximize the worst-case secrecy rate, thereby establishing a sensing-to-communication design.
The main contributions of this paper are summarized as follows:
\begin{itemize}
	
	\item 
    A hybrid RIS-assisted ISAC secure transmission architecture is proposed, in which the RIS panel is partitioned into a reflection region and a sensing reception region to physically separate communication reflection from sensing reception. The 3D position of Eve is estimated from the joint observations collected by the BS and sensing reception region, and the corresponding localization CRLB is derived. Based on the CRLB, a location uncertainty ellipsoid associated with the sensing accuracy is further constructed, providing a sensing-assisted uncertainty model for robust secure transmission design.
	
	\item 
	A two-stage optimization framework is developed for sensing and secure transmission. In the sensing stage, the relationship between the CRLB and the equivalent illumination energy is exploited to formulate the sensing optimization, which is solved by jointly optimizing the sensing beamformer and RIS phase shifts using the closed-form maximum-ratio transmission (MRT) solution and Riemannian conjugate gradient (RCG) method. In the communication stage, the semi-infinite secrecy constraint under uncertain Eve location is transformed into a linear matrix inequality (LMI) by using first-order Taylor expansion and the S-lemma. The BS beamforming is optimized via semidefinite relaxation (SDR), while the RIS phase-shift subproblem is solved using a logarithmic barrier function and RCG, thereby achieving robust maximization of the worst-case secrecy rate. 
	
	\item 
	Numerical simulations are conducted to validate the proposed architecture and optimization methods. The relationship between the CRLB and location uncertainty ellipsoid is investigated, and the sensing optimization is shown to improve localization accuracy. The proposed scheme is compared with benchmarks with perfect Eve location information, without RIS, and with random RIS phases under different BS transmit powers and RIS sizes. The results demonstrate robust secrecy performance under Eve location uncertainty and reveal the tradeoff between sensing accuracy and robust secrecy performance.
	
\end{itemize}

The remainder of this paper is organized as follows. The system model is introduced in Section II. In Section III, the problem formulation is presented and the proposed optimization algorithm is developed. Numerical results are provided in Section IV, followed by the conclusion in Section V.

\subsubsection*{Notation}
\addcontentsline{toc}{section}{Notation}

Scalars, vectors, and matrices are denoted by italic letters, boldface lowercase letters, and boldface uppercase letters, respectively. The following notations are used throughout this paper:
$(\cdot)^T$, $(\cdot)^*$, and $(\cdot)^H$ denote the transpose, conjugate, and conjugate transpose, respectively.
$|\cdot|$ denotes the absolute value of a scalar.
$\|\cdot\|$ denotes the Euclidean norm of a vector.
$\det(\cdot)$ and $\tr(\cdot)$ denote the determinant and trace, respectively.
$\Re(\cdot)$ and $\Im(\cdot)$ denote the real and imaginary parts, respectively.
$\operatorname{diag}(\cdot)$ denotes the diagonalization operator.
$\odot$ and $\otimes$ denote the Hadamard and Kronecker products, respectively.
$\mathbb{E}[\cdot]$ denotes statistical expectation.
$\mathcal{CN}(\boldsymbol{\mu},\boldsymbol{\Sigma})$ denotes a complex Gaussian distribution with mean $\boldsymbol{\mu}$ and covariance matrix $\boldsymbol{\Sigma}$.
$\mathbf{I}_N$ denotes the $N\times N$ identity matrix.
$\succeq$ and $\succ$ denote positive semidefinite and positive definite ordering, respectively.
$\lfloor\cdot\rfloor$ denotes the floor operation.
$j=\sqrt{-1}$ denotes the imaginary unit and
$\mathcal{O}(\cdot)$ denotes the computational complexity order.

\section{System Model}
\begin{figure}[htbp]
	\centering
	\includegraphics[width=0.4\textwidth]{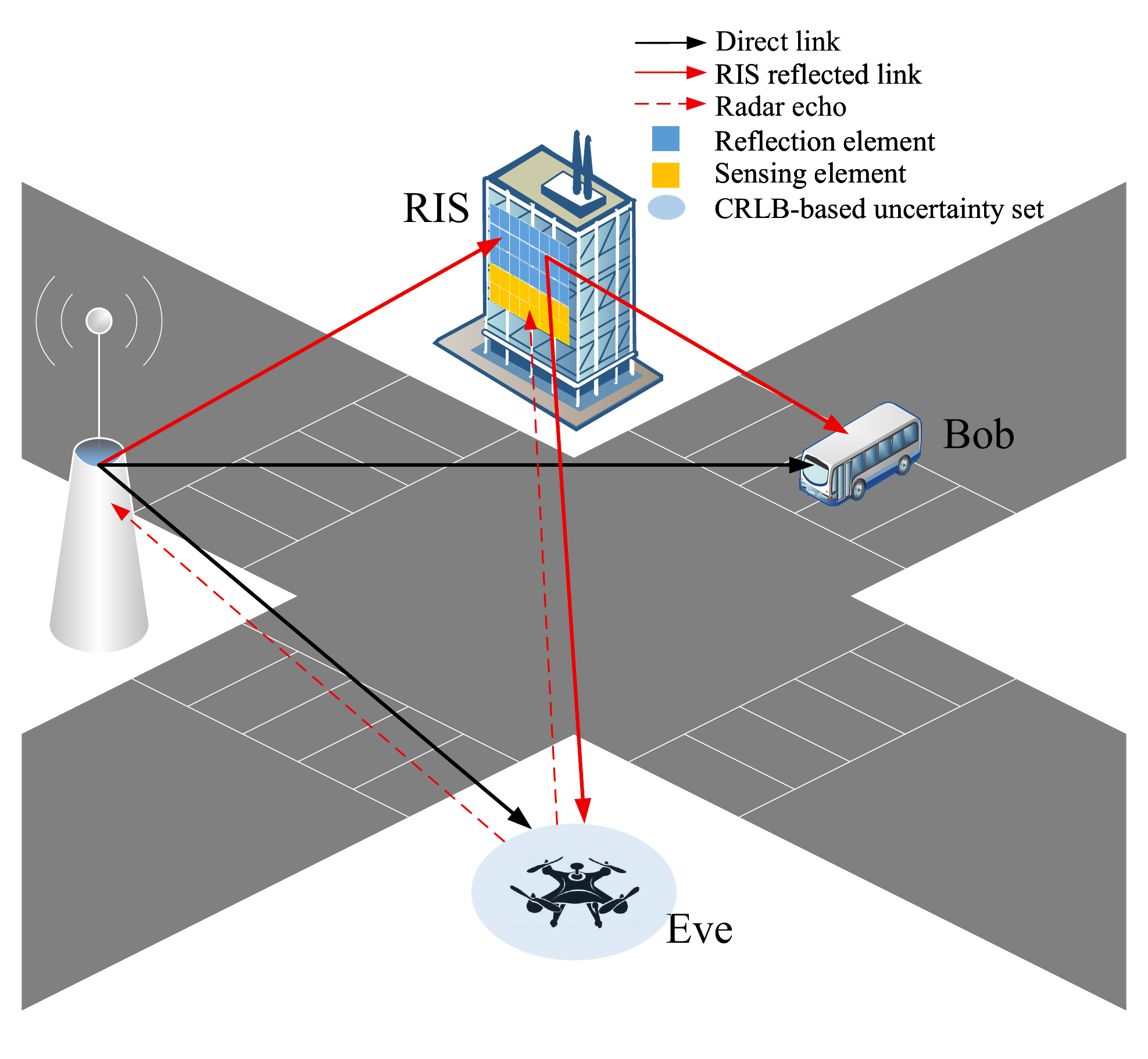}
	\caption{RIS-assisted secure ISAC transmission system model.}
	\label{fig1}
\end{figure}

As illustrated in Fig.~\ref{fig1}, we consider a 3D hybrid RIS-assisted secure ISAC
system consisting of a BS, a hybrid RIS, a single-antenna legitimate user (Bob), and a single-antenna unmanned aerial vehicle (UAV) acting as Eve, with direct links among the BS, RIS, Bob, and Eve. 
The azimuth and zenith angles are denoted by $\phi$ and $\theta$, respectively. Let $d$ and $\lambda$ denote the antenna spacing and carrier wavelength. The corresponding wavenumber components are
$\psi_x=\frac{2\pi d}{\lambda}\sin\theta\cos\phi$ and
$\psi_y=\frac{2\pi d}{\lambda}\sin\theta\sin\phi$ \cite{vanTrees2002optimum,Bhandari2026}.
The BS employs a UPA with $M_t=M_x \times M_y$ antennas, where
$M_x$ and $M_y$ denote the numbers of elements along the
$x$- and $y$-axes, respectively. Its array response is
given by
$\mathbf a_{\mathrm{BS}}(\phi,\theta)=\mathbf a_{M_x}(\psi_x)\otimes\mathbf a_{M_y}(\psi_y)$,
where $\mathbf a_K(\psi)=[1,e^{j\psi},\ldots,e^{j(K-1)\psi}]^T$.
The hybrid RIS comprises $N=N_x\times N_y$ elements and is divided into a reflection region $\mathcal R$ and a sensing reception region $\mathcal S$. A metallic isolation strip suppresses electromagnetic coupling between the two regions, and the residual leakage is neglected \cite{Mak2008,Li2025}. The reflection region occupies the upper $N_{Ry}$ rows and contains $N_R=N_xN_{Ry}$ passive reflecting elements, with $\mathbf a_{\mathcal R}(\phi,\theta)
=\mathbf a_{N_x}(\psi_x)\otimes\mathbf a_{N_{Ry}}(\psi_y)$.
The sensing reception region contains $N_S=N_xN_{Sy}$ elements, where $N_{Sy}=N_y-N_{Ry}$, and its array response is
$\mathbf a_{\mathcal S}(\phi,\theta)=\mathbf a_{N_x}(\psi_x)\otimes
\left(e^{jN_{Ry}\psi_y}\mathbf a_{N_{Sy}}(\psi_y)\right)$,
where the phase factor accounts for its displacement relative to the RIS reference point.

\subsection{Channel Model}

All channels are modeled as Rician fading channels, which are
widely used for RIS-assisted systems with line-of-sight (LoS)
components~\cite{wu2021,Wu2019}. Let
$\ell_{ij}=d_{ij}^{-\beta}$ denote the distance-dependent
large-scale power gain of link $i$--$j$. The corresponding
complex LoS coefficient is defined as
$\alpha_{ij}=\sqrt{\ell_{ij}}e^{-j\frac{2\pi}{\lambda}d_{ij}}$,
where $d_{ij}$ denotes the link distance and $\beta$ is the
path-loss exponent. The same large-scale power gain is applied
to both LoS and NLoS components.
The channel matrix from the BS to the reflection region $\mathcal R$ is modeled as
\begin{equation}
	\mathbf{H}_{br}^{\mathcal{R}}
	=
	\sqrt{\frac{\mathcal{K}_{br}}{\mathcal{K}_{br}+1}}
	\alpha_{br}
	\mathbf{H}_{br,\mathrm{LOS}}^{\mathcal{R}}
	+
	\sqrt{\frac{\ell_{br}}{\mathcal{K}_{br}+1}}
	\mathbf{H}_{br,\mathrm{NLOS}}^{\mathcal{R}},
	\label{eq:Hbr}
\end{equation}
where $\mathbf{H}_{br,\mathrm{LOS}}^{\mathcal{R}}=\mathbf{a}_{\mathcal{R}}(\phi_{r,b},\theta_{r,b})\mathbf{a}_{\mathrm{BS}}^{H}(\phi_{b,r},\theta_{b,r})$, and the NLoS component $\mathbf{H}_{br,\mathrm{NLOS}}^{\mathcal{R}}$ consists of independent and identically distributed (i.i.d.) entries following $\mathcal{CN}(0,1)$. According to electromagnetic reciprocity, the reverse channel is given by $\mathbf{H}_{rb}^{\mathcal{R}}=(\mathbf{H}_{br}^{\mathcal{R}})^T$. The direct leakage channel from the BS to the sensing reception region $\mathcal S$ is modeled as
\begin{equation}
	\mathbf{H}_{bS}
	=
	\sqrt{\frac{\mathcal{K}_{bS}}{\mathcal{K}_{bS}+1}}
	\alpha_{bS}
	\mathbf{H}_{bS,\mathrm{LOS}}
	+
	\sqrt{\frac{\ell_{bS}}{\mathcal{K}_{bS}+1}}
	\mathbf{H}_{bS,\mathrm{NLOS}},
	\label{eq:HbS}
\end{equation}
where $\mathbf{H}_{bS,\mathrm{LOS}}=\mathbf{a}_{\mathcal{S}}(\phi_{S,b},\theta_{S,b})\mathbf{a}_{\mathrm{BS}}^{H}(\phi_{b,S},\theta_{b,S})$. For a generic node $k\in\{u,e\}$, where $u$ and $e$ denote Bob and Eve, respectively, the channels from the reflection region $\mathcal R$ to node $k$ and from the BS to node $k$ are respectively modeled as
\begin{align}
	\mathbf{h}_{rk}^{\mathcal{R}}
	&=
	\sqrt{\frac{\mathcal{K}_{rk}}{\mathcal{K}_{rk}+1}}
	\alpha_{rk}^*
	\mathbf{a}_{\mathcal{R}}(\phi_{r,k},\theta_{r,k})
	+
	\sqrt{\frac{\ell_{rk}}{\mathcal{K}_{rk}+1}}
	\mathbf{h}_{rk,\mathrm{NLOS}}^{\mathcal{R}},
	\label{eq:hrk}
	\\
	\mathbf{h}_{bk}
	&=
	\sqrt{\frac{\mathcal{K}_{bk}}{\mathcal{K}_{bk}+1}}
	\alpha_{bk}^*
	\mathbf{a}_{\mathrm{BS}}(\phi_{b,k},\theta_{b,k})
	+
	\sqrt{\frac{\ell_{bk}}{\mathcal{K}_{bk}+1}}
	\mathbf{h}_{bk,\mathrm{NLOS}}.
	\label{eq:hbk}
\end{align}
The channel from the sensing reception region $\mathcal S$ to Eve is modeled as
\begin{equation}
	\mathbf{h}_{Se}
	=
	\sqrt{\frac{\mathcal{K}_{Se}}{\mathcal{K}_{Se}+1}}
	\alpha_{Se}^*
	\mathbf{a}_{\mathcal{S}}(\phi_{S,e},\theta_{S,e})
	+
	\sqrt{\frac{\ell_{Se}}{\mathcal{K}_{Se}+1}}
	\mathbf{h}_{Se,\mathrm{NLOS}}.
	\label{eq:hSe}
\end{equation}

In the above expressions, $\mathcal{K}_{(\cdot)}$ denotes the Rician factor of the corresponding link. All NLoS components consist of independent complex Gaussian entries following $\mathcal{CN}(0,1)$ before large-scale path-loss scaling.

\subsection{Sensing Model}

The system operates in a TDD mode, where each frame has a duration of $T_{\mathrm{frame}}$. The sensing duration ratio is denoted by $\tau\in(0,1)$, yielding $L=\left\lfloor \tau T_{\mathrm{frame}}f_s\right\rfloor$ snapshots, where $f_s$ is the sampling frequency.
Eve is modeled as a Swerling-I point scatterer with radar cross section (RCS) $\sigma_E\sim\operatorname{Exp}(\bar{\sigma})$, which captures target fluctuation characteristics~\cite{Geiger2026}. The corresponding scattering coefficient is given by $\alpha_e=\sqrt{\sigma_E}e^{j\phi_\sigma}$, where $\phi_\sigma\sim\operatorname{Uniform}[0,2\pi)$.
Eve is assumed to follow a constant-velocity motion model. However, within each short frame, its position is treated as constant and is updated only between adjacent frames. The scattering coefficient $\alpha_e$ is also assumed to remain constant over the $L$ snapshots within each sensing frame, while it may vary between adjacent frames.

\subsubsection{Signal Model for the Sensing Phase}

During the sensing phase, the BS transmits sensing signals as
\begin{equation}
	\mathbf{x}_s[l]=\mathbf{w}_s d_p[l],\quad l=1,\dots,L,
\end{equation}
where $\mathbf{w}_s\in\mathbb{C}^{M_t\times 1}$ is the sensing
beamformer satisfying $\|\mathbf{w}_s\|^2\leq P_s$, with
$P_s$ denoting the BS transmit power budget for sensing, and
$d_p[l]$ denotes the unit-modulus sensing waveform, i.e.,
$|d_p[l]|=1$.
The RIS reflection region $\mathcal R$ adopts the sensing phase-shift matrix $\boldsymbol{\Phi}_s=\operatorname{diag}(\phi_{s,1},\dots,\phi_{s,N_R})$, where $|\phi_{s,n}|=1$.
The sensing signal reaches Eve through the direct and RIS-reflected links, yielding
\begin{equation}
	s_{\mathrm{in}}[l]=\left(\mathbf{h}_{be}^{H}+\mathbf{h}_{re}^{\mathcal{R}H}\boldsymbol{\Phi}_s\mathbf{H}_{br}^{\mathcal{R}}\right)\mathbf{w}_s d_p[l]\triangleq z d_p[l],
	\label{eq:incident_signal}
\end{equation}
where $z=(\mathbf{h}_{be}^{H}+\mathbf{h}_{re}^{\mathcal{R}H}\boldsymbol{\Phi}_s\mathbf{H}_{br}^{\mathcal{R}})\mathbf{w}_s$ denotes the equivalent illumination channel. The corresponding illumination energy is $E_s=\sum_{l=1}^{L}|s_{\mathrm{in}}[l]|^2=L|z|^2$.
Eve scatters the incident signal, and the echoes are received by the BS and the sensing reception region $\mathcal S$ of the RIS.
For the BS receiver, assuming ideal self-interference cancellation~\cite{Lee2014}, the received signal is
\begin{equation}
	\tilde{\mathbf{y}}_{b}[l]=\alpha_e\mathbf{h}_{be}^*s_{\mathrm{in}}[l]+\mathbf{n}_{b}[l],
	\label{eq:yb_received}
\end{equation}
where $\mathbf{n}_{b}[l]\sim\mathcal{CN}(\mathbf{0},\sigma_b^2\mathbf{I}_{M_t})$.
For the RIS sensing reception region, the BS-to-RIS leakage $\mathbf{H}_{bS}\mathbf{x}_s[l]$ is removed through offline calibration or channel reciprocity-based estimation \cite{wu2021}. The effective received signal is therefore expressed as
\begin{equation}
	\tilde{\mathbf{y}}_{\mathcal{S}}[l]=\alpha_e\mathbf{h}_{Se}^*s_{\mathrm{in}}[l]+\mathbf{n}_{\mathcal{S}}[l],
	\label{eq:yS_received}
\end{equation}
where $\mathbf{n}_{\mathcal{S}}[l]\sim\mathcal{CN}(\mathbf{0},\sigma_{\mathcal{S}}^2\mathbf{I}_{N_S})$ denotes the noise at the RIS sensing receiver.

\subsubsection{Maximum Likelihood Angle Estimation}

For tractable angle estimation and CRLB analysis, a dominant-LoS single-source approximation is adopted, yielding $\mathbf{h}_{be}^*\approx\xi_b\mathbf{a}_{\mathrm{BS}}^*(\phi_{b,e},\theta_{b,e})$ and $\mathbf{h}_{Se}^*\approx\xi_{\mathcal S}\mathbf{a}_{\mathcal S}^*(\phi_{S,e},\theta_{S,e})$, where $\xi_b=\sqrt{\frac{\mathcal{K}_{be}}{\mathcal{K}_{be}+1}}\alpha_{be}$ and $\xi_{\mathcal S}=\sqrt{\frac{\mathcal{K}_{Se}}{\mathcal{K}_{Se}+1}}\alpha_{Se}$ denote the corresponding LoS gains. Under this nominal model, the BS and the RIS sensing reception region $\mathcal S$ independently estimate Eve's arrival angles from $L$ snapshots using the MLE, which asymptotically approaches the CRLB under standard regularity conditions and sufficiently informative observations \cite{Palm2022}.
Taking the BS as an example, the observation is modeled as $\tilde{\mathbf{y}}_b[l]\approx\mu_b\mathbf{a}_{\mathrm{BS}}^*(\phi_{b,e},\theta_{b,e})d_p[l]+\mathbf{n}_b[l]$, where $\mu_b=\alpha_e\xi_b z$ is the unknown complex amplitude. After pilot matched filtering, the correlation vector is $\mathbf r_b=\sum_{l=1}^{L}\bar d_p[l]\tilde{\mathbf y}_b[l]$. Eliminating $\mu_b$ from the likelihood function yields
\begin{equation}
	(\hat{\phi}_{b,e},\hat{\theta}_{b,e})=\arg\max_{\phi,\theta}\frac{\left|\mathbf{a}_{\mathrm{BS}}^{*}(\phi,\theta)\mathbf r_b\right|^2}{\|\mathbf{a}_{\mathrm{BS}}(\phi,\theta)\|^2}.
	\label{eq:mle_bs}
\end{equation}
Similarly, at the RIS sensing reception region, $\mu_{\mathcal S}=\alpha_e\xi_{\mathcal S}z$ and $\mathbf r_{\mathcal S}=\sum_{l=1}^{L}\bar d_p[l]\tilde{\mathbf y}_{\mathcal S}[l]$. The corresponding angle estimates are obtained as
\begin{equation}
	(\hat{\phi}_{S,e},\hat{\theta}_{S,e})=\arg\max_{\phi,\theta}\frac{\left|\mathbf{a}_{\mathcal S}^{*}(\phi,\theta)\mathbf r_{\mathcal S}\right|^2}{\|\mathbf{a}_{\mathcal S}(\phi,\theta)\|^2}.
	\label{eq:mle_ris}
\end{equation}
In numerical evaluations, the received signals follow the full Rician channel model, including NLoS components. The resulting model mismatch is captured through Monte Carlo localization errors, which are used to calibrate the uncertainty set in the subsequent robust design.

\subsubsection{Cross Localization}
Given the positions of the BS and the RIS, denoted by $\mathbf{p}_{BS}$ and $\mathbf{p}_{RIS}$, respectively, the unit direction vectors pointing toward Eve can be constructed from the estimated angles as
\begin{align}
	\mathbf{d}_b &= [\sin\hat{\theta}_{b,e}\cos\hat{\phi}_{b,e}, \sin\hat{\theta}_{b,e}\sin\hat{\phi}_{b,e}, \cos\hat{\theta}_{b,e}]^T, \\
	\mathbf{d}_{\mathcal S} &= [\sin\hat{\theta}_{S,e}\cos\hat{\phi}_{S,e}, \sin\hat{\theta}_{S,e}\sin\hat{\phi}_{S,e}, \cos\hat{\theta}_{S,e}]^T.
\end{align}
The 3D position of Eve is estimated as the midpoint of the shortest line segment connecting the two spatial rays \cite{Zhang2006}, i.e.,
\begin{equation}
	\hat{\mathbf{p}}_e = \frac{1}{2}\left[(\mathbf{p}_{BS}+\hat{t}_b\mathbf{d}_b) + (\mathbf{p}_{RIS}+\hat{t}_{\mathcal S}\mathbf{d}_{\mathcal S})\right],
	\label{eq:cross_loc}
\end{equation}
where $\hat{t}_b$ and $\hat{t}_{\mathcal S}$ are obtained by solving the least-squares minimum-distance problem between the two spatial rays.
\subsubsection{CRLB for Position Estimation}

The localization accuracy under the nominal single-source observation model is characterized by the CRLB. For the joint observations at the BS and the RIS sensing region $\mathcal S$, the unknown parameters consist of the Eve position $\mathbf{p}_e=[x_e,y_e,z_e]^T$ and two distinct equivalent complex coefficients $\mu_b$ and $\mu_{\mathcal S}$. Under the Swerling-I assumption, the scattering coefficient $\alpha_e$ is treated as constant over the $L$ snapshots within a sensing frame. Conditioned on its realization, the two equivalent complex coefficients are treated as deterministic unknown nuisance parameters. 

The full real-valued parameter vector is defined as $\boldsymbol{\vartheta}=[\Re(\mu_b),\Im(\mu_b),\Re(\mu_{\mathcal S}),\Im(\mu_{\mathcal S}),\mathbf{p}_e^T]^T\in\mathbb{R}^{7}$. Since the noise components at the BS and RIS sensing region are statistically independent, the full FIM is obtained by summing their receiver-specific contributions, embedded in the common parameter space. For compactness, the nonzero $5\times5$ block of the BS-side FIM is expressed using the local parameter vector $\boldsymbol{\vartheta}_b=[\Re(\mu_b),\Im(\mu_b),\mathbf{p}_e^T]^T$. Let $\mu_b=\mu_{bR}+j\mu_{bI}$. According to the Slepian--Bangs formula, this local FIM is given by
\begin{equation}
	\mathbf J_{\boldsymbol{\vartheta}_b}^{(b)}=
	\begin{bmatrix}
		\mathbf J_{\mu\mu}^{(b)} & \mathbf J_{\mu p}^{(b)}\\[2pt]
		(\mathbf J_{\mu p}^{(b)})^T & \mathbf J_{pp}^{(b)}
	\end{bmatrix},
\end{equation}
where the submatrices are given by
\begin{gather}
	\mathbf J_{\mu\mu}^{(b)} = \frac{2L}{\sigma_b^2}\|\mathbf{a}_{\mathrm{BS}}\|^2\mathbf I_2,\\
	[\mathbf J_{\mu p}^{(b)}]_{1,k} = \frac{2L}{\sigma_b^2}\Re\left\{\mu_b\mathbf{a}_{\mathrm{BS}}^{H}\frac{\partial\mathbf{a}_{\mathrm{BS}}}{\partial p_k}\right\},\\
	[\mathbf J_{\mu p}^{(b)}]_{2,k} = \frac{2L}{\sigma_b^2}\Im\left\{\mu_b\mathbf{a}_{\mathrm{BS}}^{H}\frac{\partial\mathbf{a}_{\mathrm{BS}}}{\partial p_k}\right\},\\
	[\mathbf J_{pp}^{(b)}]_{k,l} = \frac{2|\mu_b|^2L}{\sigma_b^2}\Re\left\{\left(\frac{\partial\mathbf{a}_{\mathrm{BS}}}{\partial p_k}\right)^H\frac{\partial\mathbf{a}_{\mathrm{BS}}}{\partial p_l}\right\}.
\end{gather}
The FIM contribution of the RIS sensing reception region $\mathcal S$ is derived similarly using the local parameter vector $\boldsymbol{\vartheta}_{\mathcal S}=[\Re(\mu_{\mathcal S}),\Im(\mu_{\mathcal S}),\mathbf{p}_e^T]^T$. By eliminating the two unknown complex coefficients through the Schur complement, the equivalent position FIM is obtained as
$\mathbf J_{\mathbf p_e}=\mathbf J_{pp}^{(b)}+\mathbf J_{pp}^{(\mathcal S)}-\mathbf J_{p\mu}^{(b)}(\mathbf J_{\mu\mu}^{(b)})^{-1}(\mathbf J_{p\mu}^{(b)})^T-\mathbf J_{p\mu}^{(\mathcal S)}(\mathbf J_{\mu\mu}^{(\mathcal S)})^{-1}(\mathbf J_{p\mu}^{(\mathcal S)})^T$.
After simplification, the position FIM can be written as
\begin{equation}
	\mathbf J_{\mathbf p_e}=\frac{2|\mu_b|^2L}{\sigma_b^2}\mathbf F_b+\frac{2|\mu_{\mathcal S}|^2L}{\sigma_{\mathcal S}^2}\mathbf F_{\mathcal S},
	\label{eq:FIM_position_direct}
\end{equation}
where
\begin{gather}
	\mathbf{F}_b = \Re\left\{\left(\frac{\partial \mathbf{a}_{\mathrm{BS}}}{\partial \mathbf{p}_e^T}\right)^H \mathbf{P}_{\mathbf{a}_{\mathrm{BS}}}^\perp \left(\frac{\partial \mathbf{a}_{\mathrm{BS}}}{\partial \mathbf{p}_e^T}\right)\right\},\\
	\mathbf{F}_{\mathcal S} = \Re\left\{\left(\frac{\partial \mathbf{a}_{\mathcal S}}{\partial \mathbf{p}_e^T}\right)^H \mathbf{P}_{\mathbf{a}_{\mathcal S}}^\perp \left(\frac{\partial \mathbf{a}_{\mathcal S}}{\partial \mathbf{p}_e^T}\right)\right\},
\end{gather}
and $\mathbf{P}_{\mathbf a}^{\perp}=\mathbf I-\frac{\mathbf a\mathbf a^H}{\|\mathbf a\|^2}$ denotes the orthogonal projection matrix. The position derivatives of the steering vectors are obtained through the chain rule as
$\frac{\partial \mathbf{a}}{\partial \mathbf{p}_e^T}=\frac{\partial \mathbf{a}}{\partial \phi}\frac{\partial \phi}{\partial \mathbf{p}_e^T}+\frac{\partial \mathbf{a}}{\partial \theta}\frac{\partial \theta}{\partial \mathbf{p}_e^T}$,
where the first-order derivatives with respect to the angles are determined by the array geometry, while the geometric derivatives are obtained from the relative position between the array and Eve.
The position CRLB under the adopted observation model is finally given by $\operatorname{CRLB}_{\mathbf p_e}=\mathbf J_{\mathbf p_e}^{-1}$. This CRLB is a nominal model-based benchmark and is not, by itself, a guaranteed covariance upper bound for the localization error produced by the estimator under the full Rician channel model. The discrepancy is evaluated using Monte Carlo localization errors when constructing and calibrating the uncertainty set for robust transmission.

\subsection{Communication Model}

During the communication phase with duration $(1-\tau)T_{\mathrm{frame}}$, the sensing region $\mathcal S$ remains inactive, while the reflection region $\mathcal R$ is configured with a communication-oriented phase-shift matrix $\boldsymbol{\Phi}_c=\operatorname{diag}(\phi_{c,1},\dots,\phi_{c,N_R})$, where $|\phi_{c,n}|=1$. The BS transmits the confidential data stream $\mathbf{x}_c=\mathbf{w}_c s$, where $\mathbb{E}[|s|^2]=1$ and $\|\mathbf{w}_c\|^2\leq P_c$, with $P_c$ denoting the communication transmit power budget.
The equivalent channel vectors of Bob and Eve are respectively defined as
\begin{gather}
	\mathbf{h}_u^H \triangleq \mathbf{h}_{bu}^H + \mathbf{h}_{ru}^{\mathcal{R}H} \boldsymbol{\Phi}_c \mathbf{H}_{br}^{\mathcal{R}},
	\label{eq:h_u}
	\\
	\mathbf{h}_e^H(\mathbf{p}_e) \triangleq \mathbf{h}_{be}^H(\mathbf{p}_e) + \mathbf{h}_{re}^{\mathcal{R}H}(\mathbf{p}_e) \boldsymbol{\Phi}_c \mathbf{H}_{br}^{\mathcal{R}}.
	\label{eq:h_e}
\end{gather}
The received signals at Bob and Eve are respectively given by $y_u=\mathbf{h}_u^H\mathbf{x}_c+n_u$ and $y_e=\mathbf{h}_e^H(\mathbf{p}_e)\mathbf{x}_c+n_e$, where $n_u\sim\mathcal{CN}(0,\sigma_u^2)$ and $n_e\sim\mathcal{CN}(0,\sigma_e^2)$ are the additive complex Gaussian noises. Accordingly, the received signal-to-interference-plus-noise ratios (SINRs) at Bob and Eve are $\gamma_b=\frac{|\mathbf{h}_u^H\mathbf{w}_c|^2}{\sigma_u^2}$ and $\gamma_e(\mathbf{p}_e)=\frac{|\mathbf{h}_e^H(\mathbf{p}_e)\mathbf{w}_c|^2}{\sigma_e^2}$.
The instantaneous secrecy rate is defined as
\begin{equation}
	R_s = \left[ \log_2(1+\gamma_b) - \log_2(1+\gamma_e) \right]^+,
\end{equation}
where $[\cdot]^+$ denotes $\max(\cdot,0)$ \cite{cui2019}. The effective secrecy rate is further defined as $R_{\mathrm{eff}}=(1-\tau)R_s$.

The above system model provides the foundations for robust secure transmission design. In the following section, the nominal position CRLB is used as a localization-accuracy benchmark, while the scale of the resulting uncertainty set is calibrated using localization errors obtained from the full Rician channel model.

\section{Sensing-Assisted Robust Secure Transmission}

In this section, we develop a sensing-assisted robust secure transmission design based on the preceding system model. Specifically, we first construct a sensing-driven positioning uncertainty set for Eve and then formulate the corresponding worst-case secrecy rate maximization problem. Since the sensing design determines the positioning accuracy and thus the uncertainty region used for secure transmission, the resulting problem is highly coupled. To exploit the sensing-to-communication temporal causality, we adopt a two-stage sequential optimization framework, where the sensing variables are first optimized to determine the uncertainty set, which is then fixed for the robust secure transmission design. This framework yields a tractable suboptimal solution to the original joint optimization problem.

\subsection{Sensing-Assisted Robust Uncertainty Set}

In the $t$-th frame, the MLE-based localization procedure
provides an estimate $\hat{\mathbf p}_e$ of Eve's 3D position.
Although the MLE can achieve asymptotic efficiency under
suitable regularity conditions, its practical localization
performance may deviate from the theoretical CRLB~\cite{Pu2026}. This is
because only a fraction of the frame is allocated to sensing,
which limits the number of available snapshots, while the
mismatch between the adopted observation model and the actual
Rician channels may further degrade localization accuracy.
Therefore, the nominal CRLB is used as a shape matrix for
constructing a robust uncertainty region, rather than being
directly interpreted as the actual error covariance.
Specifically, the uncertainty ellipsoid is constructed by
introducing a scaling factor $\kappa\geq 1$ to account for
the discrepancy between the nominal CRLB and the actual
localization errors. The uncertainty region is defined as
\begin{equation}
	\mathcal U_E =
	\left\{
	\mathbf p\in\mathbb R^3 \,\middle|\,
	(\mathbf p-\hat{\mathbf p}_e)^T
	\mathbf M^{-1}
	(\mathbf p-\hat{\mathbf p}_e)\leq 1
	\right\},
	\label{eq:U_E}
\end{equation}
where
$\mathbf M=\kappa\chi^2_{3,\alpha}\mathbf C_e$,
$\mathbf C_e=\operatorname{CRLB}_{\mathbf p_e}$ denotes
the nominal position CRLB for the current frame and sensing
design, and $\chi^2_{3,\alpha}$ is the $\alpha$-quantile of
the chi-square distribution with three degrees of freedom.
The scaling factor $\kappa$ is empirically selected based on
Monte Carlo localization errors under the considered
simulation settings and is fixed in the subsequent robust
transmission design.

\subsection{Problem Formulation and Two-Stage Solution Framework}

Based on the constructed uncertainty set, the worst-case secrecy rate maximization problem is formulated as
\begin{subequations}\label{eq:P0}
	\begin{align}
		\mathcal{P}0:\quad
		\max_{\mathbf{w}_s,\mathbf{w}_c,\boldsymbol{\Phi}_s,\boldsymbol{\Phi}_c} \min_{\mathbf{p}_e\in\mathcal{U}_e(\mathbf{w}_s,\boldsymbol{\Phi}_s)} & R_s(\mathbf{w}_c,\boldsymbol{\Phi}_c,\mathbf{p}_e) \notag \\
		\text{s.t.} \quad & \|\mathbf{w}_s\|^2\le P, \label{eq:P0a} \\
		& \|\mathbf{w}_c\|^2\le P, \label{eq:P0b} \\
		& |\phi_{s,n}|=1,\; \forall n, \label{eq:P0c} \\
		& |\phi_{c,n}|=1,\; \forall n, \label{eq:P0d}
	\end{align}
\end{subequations}
where $P_s=P_c=P$.
Problem $\mathcal{P}_0$ jointly optimizes the sensing variables $(\mathbf{w}_s,\boldsymbol{\Phi}_s)$ and communication variables $(\mathbf{w}_c,\boldsymbol{\Phi}_c)$. The sensing design determines Eve's localization accuracy and the resulting uncertainty set $\mathcal{U}_e$, while the communication design maximizes the worst-case secrecy rate over this set. Problem $\mathcal{P}_0$ is highly non-convex due to the coupling between the two designs, the unit-modulus RIS constraints, and the semi-infinite robust secrecy constraint. 

To exploit the temporal causality of the sensing and communication procedure, we adopt a two-stage sequential optimization framework. In Stage I, $(\mathbf{w}_s,\boldsymbol{\Phi}_s)$ are optimized to minimize the position CRLB and construct $\mathcal{U}_e$. In Stage II, $\mathcal{U}_e$ is fixed and $(\mathbf{w}_c,\boldsymbol{\Phi}_c)$ are optimized for worst-case secrecy rate maximization. This sequential design follows the actual sensing-to-communication order and provides a tractable suboptimal solution to $\mathcal{P}_0$.

\subsection{Stage I: Sensing Optimization}

The objective of the sensing stage is to jointly optimize the sensing beamforming vector $\mathbf{w}_s$ and the RIS sensing phase-shift matrix $\boldsymbol{\Phi}_s$ under the transmit power constraint and the unit-modulus constraints of RIS elements, so as to minimize the trace of the CRLB matrix for Eve's position estimation. By adopting the trace of the position CRLB matrix as the scalar accuracy metric, the sensing optimization problem is formulated as
\begin{subequations}\label{eq:P1}
	\begin{align}
		\mathcal{P}1:\quad
		\min_{\mathbf{w}_s,\boldsymbol{\Phi}_s}\quad&
		\tr\!\left(
		\operatorname{CRLB}_{\mathbf{p}_e}
		(\mathbf{w}_s,\boldsymbol{\Phi}_s)
		\right)\\
		\mathrm{s.t.}\quad&
		\|\mathbf{w}_s\|^2\le P,\\
		&|\phi_{s,n}|=1,\quad n=1,\ldots,N_R.
	\end{align}
\end{subequations}
The above objective involves matrix inversion and trace operations, while the optimization variables are coupled through the equivalent complex scattering amplitudes, making the problem difficult to solve directly.
Assume that the BS and the RIS sensing receiver have the same noise variance, i.e., $\sigma_b^2=\sigma_{\mathcal S}^2\triangleq\sigma^2$. According to the nominal observation model, the two equivalent complex amplitudes are given by $\mu_b=\alpha_e\xi_b z$ and $\mu_{\mathcal S}=\alpha_e\xi_{\mathcal S}z$, where $\xi_b$ and $\xi_{\mathcal S}$ denote the corresponding LoS gains. Although both observations share the same illumination coefficient $z$, their equivalent complex amplitudes are generally different. Substituting these expressions into \eqref{eq:FIM_position_direct} yields
\begin{equation}
	\begin{aligned}
		\mathbf J_{\mathbf p_e}
		=2L|z|^2|\alpha_e|^2
		\left(
		\frac{|\xi_b|^2}{\sigma^2}\mathbf F_b
		+\frac{|\xi_{\mathcal S}|^2}{\sigma^2}\mathbf F_{\mathcal S}
		\right)\
		\triangleq 2L|z|^2\mathbf Q,
	\end{aligned}
\end{equation}
where
$\mathbf Q\triangleq\frac{|\alpha_e|^2}{\sigma^2}\left(
|\xi_b|^2\mathbf F_b+|\xi_{\mathcal S}|^2\mathbf F_{\mathcal S}\right)$.
For a fixed nominal Eve position and propagation environment, the matrices $\mathbf F_b$ and $\mathbf F_{\mathcal S}$, as well as the nominal LoS gains, are independent of the sensing beamformer and RIS phase shifts. The scattering coefficient is treated as a fixed realization under the conditional CRLB framework. Therefore, $\mathbf Q$ is independent of the sensing optimization variables. Provided that the combined angular observation geometry is locally identifiable and $\mathbf Q\succ0$, the position CRLB and its trace are given by
$\operatorname{CRLB}_{\mathbf p_e}=\frac{1}{2L|z|^2}\mathbf Q^{-1}$,
and
$\operatorname{tr}\left(\operatorname{CRLB}_{\mathbf p_e}\right)=\frac{\operatorname{tr}(\mathbf Q^{-1})}{2L|z|^2}$.
Consequently, minimizing the trace of the position CRLB is equivalent to maximizing the effective illumination power $|z|^2$. Therefore,  $\mathcal{P}1$ is equivalently reduced to maximizing the effective illumination power:
\begin{subequations}\label{eq:P1_equivalent}
	\begin{align}
		\mathcal{P}1.1: \max_{\mathbf w_s,\boldsymbol{\Phi}_s} & \left| \left( \mathbf h_{be}^{H} + \mathbf h_{re}^{\mathcal RH} \boldsymbol{\Phi}_s \mathbf H_{br}^{\mathcal R} \right) \mathbf w_s \right|^2 \notag \\
		\mathrm{s.t.} \quad & \|\mathbf w_s\|^2\le P, \label{eq:P1_equivalent_a} \\
		& |\phi_{s,n}|=1,\quad n=1,\ldots,N_R . \label{eq:P1_equivalent_b}
	\end{align}
\end{subequations}
Problem $\mathcal{P}$1.1 is jointly non-convex with respect to $(\mathbf w_s,\boldsymbol{\Phi}_s)$ due to the coupling between active beamforming and RIS phase shifts, as well as the unit-modulus constraints of RIS elements. To address this issue, an alternating optimization (AO) framework is adopted, where one variable is iteratively optimized while fixing the other. The AO algorithm produces a non-decreasing sequence of sensing objective values when each block is updated with a monotonic line search. Since the objective is upper bounded under the transmit-power constraint, the resulting objective sequence converges. Under standard regularity conditions, any limit point of the AO iterates satisfies the first-order stationarity conditions of $\mathcal{P}1.1$.

\subsubsection{Subproblem 1: Optimal $\mathbf w_s$ with Fixed $\boldsymbol{\Phi}_s$}
With $\boldsymbol{\Phi}_s$ fixed, define the equivalent channel vector as $\mathbf h_{\mathrm{eff}}^H\triangleq\mathbf h_{be}^H+\mathbf h_{re}^{\mathcal RH}\boldsymbol{\Phi}_s\mathbf H_{br}^{\mathcal R}\in\mathbb C^{1\times M_t}$. Then, the objective function of $\mathcal{P}$1.1 reduces to $f(\mathbf w_s)=|\mathbf h_{\mathrm{eff}}^H\mathbf w_s|^2$. By applying the Cauchy--Schwarz inequality, we have $|\mathbf h_{\mathrm{eff}}^H\mathbf w_s|\le\|\mathbf h_{\mathrm{eff}}\|\|\mathbf w_s\|\le\sqrt{P}\|\mathbf h_{\mathrm{eff}}\|$. The equality condition is achieved when the transmit beamforming vector is aligned with the equivalent channel vector and the available transmit power is fully utilized. Therefore, the global optimal solution is obtained by MRT principle, given by
$\mathbf w_s^\star = \sqrt{P} \frac{\mathbf h_{\mathrm{eff}}}{\|\mathbf h_{\mathrm{eff}}\|}
$.

\subsubsection{Subproblem 2: $\boldsymbol{\Phi}_s$ Optimization with Fixed $\mathbf w_s$}

With a fixed sensing beamformer $\mathbf w_s$, define $\boldsymbol{\phi}_s=[\phi_{s,1},\ldots,\phi_{s,N_R}]^T$ and $\boldsymbol{\Phi}_s=\operatorname{diag}(\boldsymbol{\phi}_s)$. Let $c_0=\mathbf h_{be}^{H}\mathbf w_s$ and $\mathbf v=\operatorname{diag}(\mathbf h_{re}^{\mathcal R *})\mathbf H_{br}^{\mathcal R}\mathbf w_s$. Then, $\left|\mathbf h_{\mathrm{eff}}^{H}\mathbf w_s\right|^2=\left|c_0+\boldsymbol{\phi}_s^{T}\mathbf v\right|^2$, and the phase optimization problem becomes
\begin{align*}
	\mathcal{P}1.2:\quad
	\min_{\boldsymbol{\phi}_s\in\mathcal M_s}\quad
	&g(\boldsymbol{\phi}_s)=-\left|c_0+\boldsymbol{\phi}_s^{T}\mathbf v\right|^2,
\end{align*}
where $\mathcal M_s=\left\{\boldsymbol{\phi}_s\in\mathbb C^{N_R}\mid|\phi_{s,n}|=1,\;n=1,\ldots,N_R\right\}$ is the complex circle product manifold.
We solve $\mathcal{P}1.2$ using the RCG algorithm. Specifically, the Euclidean gradient is projected onto the tangent space to obtain the Riemannian gradient, followed by a conjugate search direction and retraction-based update that preserves the unit-modulus constraints. The gradient derivation is given in Appendix~\ref{app:gradient_sensing}. At the $j$-th iteration, $\boldsymbol{\phi}_s^{(j+1)}=R_{\boldsymbol{\phi}_s^{(j)}}\left(\alpha_j\mathbf d^{(j)}\right)$, where $\alpha_j$ is determined by Armijo line search and $\mathbf d^{(j)}$ is the RCG search direction. The iteration stops when $\left|\operatorname{grad} g(\boldsymbol{\phi}_s^{(j)})\right|<\epsilon_{\mathrm{in}}$ or the maximum number of iterations is reached.

\subsection{Stage II: Robust Secure Transmission Design}

During the communication slot, the objective is to maximize the instantaneous secrecy rate for the worst-case Eve position, while satisfying the transmit power constraint at the BS and the unit-modulus constraint of the RIS reflection coefficients. Let
$
	R_s(\mathbf{w}_c,\boldsymbol{\Phi}_c,\mathbf{p}_e)
	=
	\left[
	\log_2\bigl(1+\gamma_b(\mathbf{w}_c,\boldsymbol{\Phi}_c)\bigr)
	-
	\log_2\bigl(1+\gamma_e(\mathbf{w}_c,\boldsymbol{\Phi}_c,\mathbf{p}_e)\bigr)
	\right]^{+}
$.
The robust optimization problem is formulated as
\begin{subequations}\label{eq:P2}
	\begin{align}
		\mathcal{P}2: \max_{\mathbf{w}_c,\boldsymbol{\Phi}_c} & \min_{\mathbf{p}_e\in\mathcal{U}_e} R_s(\mathbf{w}_c,\boldsymbol{\Phi}_c,\mathbf{p}_e) \notag \\
		\mathrm{s.t.} \quad & \|\mathbf{w}_c\|^2\leq P, \label{eq:P2_a} \\
		& |\phi_{c,n}|=1,\quad n=1,\ldots,N_R . \label{eq:P2_b}
	\end{align}
\end{subequations}

The above problem exhibits a max--min structure, and the uncertainty set \(\mathcal{U}_e\) is a continuous ellipsoid. Therefore, it belongs to a typical class of semi-infinite optimization problems. We focus on the nontrivial regime with a positive secrecy-rate target $\Gamma>0$. When no positive secrecy rate is feasible, the optimal secrecy rate is zero.
In this regime, the operator \([\cdot]^{+}\) does not affect the epigraph constraint, and \(R_s\geq\Gamma\) is equivalent to
$\log_2\bigl(1+\gamma_b\bigr)-\log_2\bigl(1+\gamma_e\bigr)\geq\Gamma$.
By introducing the auxiliary variables \(\Gamma\) and \(\eta\geq0\), problem \(\mathcal{P}2\) can be equivalently represented in the positive-secrecy-rate regime by the following epigraph formulation:
\begin{subequations}\label{eq:P2_epi}
	\begin{align}
		\mathcal{P}2.1: \max_{\mathbf{w}_c,\boldsymbol{\Phi}_c,\Gamma,\eta} & \quad \Gamma \notag \\
		\mathrm{s.t.} \quad & \gamma_b(\mathbf{w}_c,\boldsymbol{\Phi}_c)\geq\eta, \label{eq:P2_epi_a} \\
		& \max_{\mathbf{p}_e\in\mathcal{U}_e}\gamma_e(\mathbf{p}_e)\leq\bar{\gamma}_e(\eta,\Gamma), \label{eq:P2_epi_b} \\
		& \|\mathbf{w}_c\|^2\leq P, \label{eq:P2_epi_c} \\
		& |\phi_{c,n}|=1,\quad n=1,\ldots,N_R, \label{eq:P2_epi_d} \\
		& \Gamma\geq0,\quad \eta\geq0. \label{eq:P2_epi_e}
	\end{align}
\end{subequations}
where \(\eta\) is an auxiliary variable representing a lower bound on the Bob's SINR, and $\bar{\gamma}_e(\eta,\Gamma)=\frac{1+\eta}{2^{\Gamma}}-1$
denotes the corresponding maximum admissible Eve SINR threshold. The derivation follows directly from $R_s\geq\Gamma$, which gives $\log_2\frac{1+\gamma_b}{1+\gamma_e}\geq\Gamma$, and hence $\gamma_e\leq\frac{1+\gamma_b}{2^\Gamma}-1$.
Introducing the auxiliary variable \(\eta\leq\gamma_b\) gives the sufficient condition
$\gamma_e\leq\frac{1+\eta}{2^{\Gamma}}-1$.
Since $\eta$ is upper bounded by $\gamma_b$, while increasing $\eta$ enlarges the admissible Eve-SINR threshold $\bar{\gamma}_e(\eta,\Gamma)$, the largest feasible value of $\eta$ is optimal, yielding $\eta=\gamma_b$. Hence, the reformulation is equivalent at the optimum. The main challenge of this problem lies in the robust constraint in \eqref{eq:P2_epi}, which is a semi-infinite constraint requiring the Eve's SINR to remain below the prescribed threshold for all possible locations within the continuous uncertainty region.
Let \(\mathbf{x}=\mathbf{p}_e-\hat{\mathbf{p}}_e\). Then, the uncertainty region can be equivalently represented by the position-error set
$\mathcal{X}_e=\left\{\mathbf{x}\in\mathbb{R}^3\mid
\mathbf{x}^{T}\mathbf{M}^{-1}\mathbf{x}\leq1\right\}$.
When the sensing stage yields a sufficiently small localization uncertainty region, the channel variation over the corresponding uncertainty region can be accurately approximated by a first-order Taylor expansion. Accordingly, we have
\begin{equation}
	\mathbf{h}_e(\hat{\mathbf{p}}_e+\mathbf{x};\boldsymbol{\Phi}_c)
	\approx
	\mathbf{h}_0(\boldsymbol{\Phi}_c)
	+
	\mathbf{J}_h(\boldsymbol{\Phi}_c)\mathbf{x},
\end{equation}
where
$\mathbf{h}_0(\boldsymbol{\Phi}_c)=\mathbf{h}_e(\hat{\mathbf{p}}_e;\boldsymbol{\Phi}_c)$, 
$\mathbf{J}_h(\boldsymbol{\Phi}_c)=\left.\frac{\partial\mathbf{h}_e(\mathbf{p}_e;\boldsymbol{\Phi}_c)}{\partial\mathbf{p}_e^{T}}\right|_{\mathbf{p}_e=\hat{\mathbf{p}}_e}$
denote the Eve channel evaluated at the estimated position and the Jacobian matrix of the Eve channel vector with respect to the Eve position, both of which depend on the RIS phase-shift matrix \(\boldsymbol{\Phi}_c\).
Let \(\mathbf{W}=\mathbf{w}_c\mathbf{w}_c^{H}\), where \(\mathbf{W}\succeq0\) and \(\operatorname{rank}(\mathbf{W})=1\). Then the Eve SINR can be rewritten as a real-valued quadratic function with respect to the position error vector \(\mathbf{x}\):
$\gamma_e(\mathbf{x})=\mathbf{x}^{T}\mathbf{A}\mathbf{x}+2\mathbf{b}^{T}\mathbf{x}+c$,
where
$\mathbf{A}	=\frac{1}{\sigma_e^2}\Re\!\left(\mathbf{J}_h^{H}\mathbf{W}\mathbf{J}_h\right)$, 
$\mathbf{b}=\frac{1}{\sigma_e^2}\Re\!\left(\mathbf{J}_h^{H}\mathbf{W}\mathbf{h}_0\right)$,
$c=\frac{1}{\sigma_e^2}\mathbf{h}_0^{H}\mathbf{W}\mathbf{h}_0$.
Since \(\mathbf{W}\succeq0\), the matrix \(\mathbf{A}\) is real symmetric and positive semidefinite. Therefore, the robust SINR constraint requires the following quadratic inequality to be satisfied for all possible position errors within the uncertainty ellipsoid
$\bar{\gamma}_e-\mathbf{x}^{T}\mathbf{A}\mathbf{x}-2\mathbf{b}^{T}\mathbf{x}-c\geq0,\forall \mathbf{x}\in\mathcal{X}_e $.
According to the S-lemma, the above implication holds if and only if there exists a scalar \(\lambda\geq0\) such that
\begin{equation}\label{eq:lmi}
	\mathbf{S}(\mathbf{W},\lambda,\Gamma,\eta) \triangleq
	\begin{bmatrix}
		\lambda\mathbf{M}^{-1}-\mathbf{A} & -\mathbf{b}\\
		-\mathbf{b}^{T} & \bar{\gamma}_e-c-\lambda
	\end{bmatrix}
	\succeq0 .
\end{equation}
Since \(\mathbf{M}\succ0\), the uncertainty set is a compact ellipsoid with a nonempty interior. In particular, \(\mathbf{x}=0\) strictly satisfies \(\mathbf{x}^{T}\mathbf{M}^{-1}\mathbf{x}=0<1\). Hence, the strict-feasibility condition required by the S-lemma is satisfied, and the robust quadratic constraint admits an exact LMI reformulation under the adopted first-order channel approximation.
Now introduce the rank-one beamforming covariance matrix \(\mathbf{W}=\mathbf{w}_c\mathbf{w}_c^{H}\), which satisfies \(\mathbf{W}\succeq0\) and \(\operatorname{rank}(\mathbf{W})=1\). By combining the above transformations, the robust secure transmission problem can be reformulated as
\begin{subequations}\label{eq:P2_lmi}
	\begin{align}
		\mathcal{P}2.2: \max_{\mathbf{W},\boldsymbol{\Phi}_c,\Gamma,\eta,\lambda} & \quad \Gamma \notag \\
		\mathrm{s.t.} \quad & \mathbf{W}\succeq0, \label{eq:P2_lmi_a} \\
		& \operatorname{rank}(\mathbf{W})=1, \label{eq:P2_lmi_b} \\
		& \operatorname{tr}(\mathbf{W})\leq P, \label{eq:P2_lmi_c} \\
		& |\phi_{c,n}|=1,\quad n=1,\ldots,N_R, \label{eq:P2_lmi_d} \\
		& \frac{1}{\sigma_u^2}\operatorname{tr}\!\left(\mathbf{h}_u\mathbf{h}_u^{H}\mathbf{W}\right)\geq\eta, \label{eq:P2_lmi_e} \\
		& \mathbf{S}(\mathbf{W},\lambda,\Gamma,\eta)\succeq0, \label{eq:P2_lmi_f} \\
		& \lambda\geq0,\quad \Gamma\geq0,\quad \eta\geq0. \label{eq:P2_lmi_g}
	\end{align}
\end{subequations}
Although the semi-infinite robust constraint is exactly reformulated as an LMI under the adopted first-order channel approximation, problem \(\mathcal{P}2.2\) remains non-convex due to the rank-one structure of \(\mathbf{W}\), the unit-modulus constraint of the RIS phase-shift coefficients, and the strong coupling between the beamforming covariance matrix \(\mathbf{W}\) and the RIS phase-shift matrix \(\boldsymbol{\Phi}_c\) in the LMI constraint. To address these challenges, an AO framework is adopted, in which the beamforming and RIS phase-shift variables are optimized alternately.

\subsubsection{Beamforming Optimization Subproblem}

With the RIS phase-shift matrix \(\boldsymbol{\Phi}_c\) fixed, the equivalent legitimate channel \(\mathbf{h}_u\), the nominal Eve channel \(\mathbf{h}_0\), and the channel Jacobian matrix \(\mathbf{J}_h\) are regarded as constants. By introducing the covariance matrix \(\mathbf{W}=\mathbf{w}_c\mathbf{w}_c^{H}\), the Bob's SINR can be expressed as \(\gamma_b=\frac{1}{\sigma_u^2}\operatorname{tr}(\mathbf{h}_u\mathbf{h}_u^{H}\mathbf{W})\). Moreover, the matrices \(\mathbf{A}\), \(\mathbf{b}\), and \(c\) in the LMI constraint are affine functions of \(\mathbf{W}\). Therefore, for a given secrecy rate target \(\Gamma_0\), the beamforming optimization problem can be transformed into a semidefinite feasibility problem by relaxing the rank-one constraint of \(\mathbf{W}\):
\begin{subequations}\label{eq:SDP_feas}
	\begin{align}
		\mathcal{P}2.2.1: \mathrm{find} \quad & \{\mathbf{W},\lambda,\eta\} \notag \\
		\mathrm{s.t.} \quad & \mathbf{W}\succeq0, \label{eq:SDP_feas_a} \\
		& \lambda\geq0,\quad \eta\geq0, \label{eq:SDP_feas_b} \\
		& \operatorname{tr}(\mathbf{W})\leq P, \label{eq:SDP_feas_c} \\
		& \frac{1}{\sigma_u^2}\operatorname{tr}\!\left(\mathbf{h}_u\mathbf{h}_u^{H}\mathbf{W}\right)\geq\eta, \label{eq:SDP_feas_d} \\
		& \mathbf{S}(\mathbf{W},\lambda,\Gamma_0,\eta)\succeq0. \label{eq:SDP_feas_e}
	\end{align}
\end{subequations}
where \(\bar{\gamma}_e=\frac{1+\eta}{2^{\Gamma_0}}-1\). The feasibility of the relaxed SDP determines whether the secrecy-rate target is achievable in the lifted covariance relaxation. A bisection search is therefore adopted to obtain the maximum secrecy-rate target \(\Gamma_{\mathrm{SDR}}^{\star}\) supported by the relaxed SDP. Specifically, if the relaxed SDP is feasible, \(\Gamma_0\) is increased. Otherwise, it is decreased. If the optimal covariance matrix is rank one, the corresponding beamforming vector can be directly recovered. Otherwise, Gaussian randomization is employed to generate candidate rank-one beamforming vectors. Only candidates satisfying the original power and robust secrecy constraints are retained, and the feasible candidate yielding the highest secrecy rate is selected. If no randomized candidate is feasible, the incumbent feasible beamforming solution is retained.

\subsubsection{RIS Phase-Shift Optimization Subproblem}

With the recovered rank-one beamforming vector \(\mathbf{w}_c\) fixed, equivalently with \(\mathbf{W}=\mathbf{w}_c\mathbf{w}_c^{H}\) fixed, the RIS phase-shift optimization subproblem can be formulated as
\begin{subequations}\label{eq:phase_sub}
	\begin{align}
		\max_{\boldsymbol{\phi}_c,\Gamma,\eta,\lambda} \quad & \Gamma \notag \\
		\mathrm{s.t.} \quad & \boldsymbol{\phi}_c\in\mathcal{M}, \label{eq:phase_sub_a} \\
		& \Gamma\geq0,\quad \lambda\geq0,\quad \eta\geq0, \label{eq:phase_sub_b} \\
		& \frac{1}{\sigma_u^2}|\mathbf{h}_u^{H}(\boldsymbol{\phi}_c)\mathbf{w}_c|^2\geq\eta, \label{eq:phase_sub_c} \\
		& \mathbf{S}(\boldsymbol{\phi}_c,\lambda,\Gamma,\eta)\succeq\mathbf{0}. \label{eq:phase_sub_d}
	\end{align}
\end{subequations}
where
\begin{equation}
	\mathbf{S}(\boldsymbol{\phi}_c,\lambda,\Gamma,\eta) \triangleq
	\begin{bmatrix}
		\lambda\mathbf{M}^{-1} - \mathbf{A}(\boldsymbol{\phi}_c) & -\mathbf{b}(\boldsymbol{\phi}_c) \\
		-\mathbf{b}^{T}(\boldsymbol{\phi}_c) & \bar{\gamma}_e(\eta,\Gamma) - c(\boldsymbol{\phi}_c) - \lambda
	\end{bmatrix},
\end{equation}
and $\bar{\gamma}_e(\eta,\Gamma)=\frac{1+\eta}{2^{\Gamma}}-1$.
Here, \(\mathbf{A}(\boldsymbol{\phi}_c)\), \(\mathbf{b}(\boldsymbol{\phi}_c)\), and \(c(\boldsymbol{\phi}_c)\) denote the equivalent quadratic-form coefficients obtained after transforming the real-valued position uncertainty variables. The feasible set is given by
$\mathcal{M}=\{\boldsymbol{\phi}_c\in\mathbb{C}^{N_R}\mid|\phi_{c,n}|=1,\, n=1,\ldots,N_R\}$,
which represents the complex circle product manifold induced by the unit-modulus constraints. This problem involves both non-convex manifold constraints and positive semidefinite matrix inequality constraints.

To avoid repeatedly solving a semidefinite program in the RIS phase-shift block, the positive-semidefinite constraint is handled using a logarithmic barrier, which is incorporated into the objective function. Specifically, the logarithmic barrier objective is defined as
\begin{equation}
	\Phi_{\mu}(\boldsymbol{\phi}_c,\lambda,\Gamma,\eta) \triangleq \Gamma + \mu \log\det \mathbf{S}(\boldsymbol{\phi}_c,\lambda,\Gamma,\eta),
\end{equation}
where \(\mu>0\) is the barrier parameter. Then, the barrier optimization problem is constructed as
\begin{subequations}\label{eq:barrier}
	\begin{align}
		\mathcal{P}2.2.2: \max_{\boldsymbol{\phi}_c,\lambda,\Gamma,\eta} \quad & \Phi_{\mu}(\boldsymbol{\phi}_c,\lambda,\Gamma,\eta) \notag \\
		\mathrm{s.t.} \quad & \boldsymbol{\phi}_c\in\mathcal{M}, \label{eq:barrier_a} \\
		& \lambda\geq0,\quad \Gamma\geq0,\quad \eta\geq0, \label{eq:barrier_b} \\
		& \frac{1}{\sigma_u^2}|\mathbf{h}_u^{H}(\boldsymbol{\phi}_c)\mathbf{w}_c|^2\geq\eta, \label{eq:barrier_c} \\
		& \mathbf{S}(\boldsymbol{\phi}_c,\lambda,\Gamma,\eta)\succ0. \label{eq:barrier_d}
	\end{align}
\end{subequations}
Since the log-det barrier is defined only on the positive-definite domain, the barrier iterations require an initial point satisfying \(\mathbf{S}\succ0\) and the remaining constraints. Such a strictly feasible point \((\boldsymbol{\phi}_c^{(0)},\lambda^{(0)},\Gamma^{(0)},\eta^{(0)})\) is obtained through a dedicated initialization procedure, which guarantees that \(\mathbf{S}(\boldsymbol{\phi}_c^{(0)},\lambda^{(0)},\Gamma^{(0)},\eta^{(0)})\succ0\) and the Bob SINR constraint are satisfied, thereby ensuring the validity of the subsequent barrier optimization process.

Based on the above barrier formulation, a two-layer iterative strategy is adopted to solve the RIS phase-shift optimization subproblem. The outer loop gradually decreases the barrier parameter \(\mu\), while the inner loop employs a cyclic block-coordinate ascent strategy to update \(\lambda\), \(\Gamma\), \(\boldsymbol{\phi}_c\), and \(\eta\). The inner-loop procedure is summarized as follows.

\paragraph{Update \(\lambda\)}
With \(\boldsymbol{\phi}_c\), \(\Gamma\), and \(\eta\) fixed, the matrices \(\mathbf{A}\), \(\mathbf{b}\), \(c\), and \(\bar{\gamma}_e\) become constants. Consequently, the LMI matrix \(\mathbf{S}\) is an affine function of \(\lambda\), which can be expressed as \(\mathbf{S}(\lambda)=\mathbf{S}_0+\lambda\mathbf{S}_{\lambda}\), where \(\mathbf{S}_0\) and \(\mathbf{S}_{\lambda}\) are constant matrices determined by the current values of \(\boldsymbol{\phi}_c\), \(\Gamma\), and \(\eta\). Accordingly, the barrier objective function with respect to \(\lambda\) is given by \(\Phi_{\mu}(\lambda)=\mu\log\det\mathbf{S}(\lambda)+\Gamma\). Using standard matrix differential identities, the first- and second-order derivatives of \(\Phi_{\mu}(\lambda)\) are obtained as
$\Phi_{\mu}'(\lambda)=\mu\operatorname{tr}\!\left(\mathbf{S}^{-1}(\lambda)\mathbf{S}_{\lambda}\right)$, 
$\Phi_{\mu}''(\lambda)=-\mu\operatorname{tr}\!\left(\mathbf{S}^{-1}(\lambda)\mathbf{S}_{\lambda}\mathbf{S}^{-1}(\lambda)\mathbf{S}_{\lambda}\right)$.
Since \(\mathbf{S}(\lambda)\succ0\), the second-order derivative is non-positive, indicating that the barrier objective is concave with respect to \(\lambda\). Since the subproblem is concave over its feasible domain, any stationary point is globally optimal, and Newton's method with backtracking is employed to obtain the maximizer. Specifically, the update rule is given by
\begin{equation}
	\lambda^{(k+1)} = \lambda^{(k)} - \frac{\Phi_{\mu}'(\lambda^{(k)})}{\Phi_{\mu}''(\lambda^{(k)})},
\end{equation}
where a backtracking line search is employed to preserve \(\lambda^{(k+1)}\ge0\) and \(\mathbf{S}(\lambda^{(k+1)})\succ0\), while satisfying the Armijo ascent condition.

\paragraph{Update \(\Gamma\)}
When \(\boldsymbol{\phi}_c\), \(\lambda\), and \(\eta\) are fixed, the barrier objective function only depends on the scalar variable \(\Gamma\), which is given by \(\Phi_\mu(\Gamma)=\Gamma+\mu\log\det\mathbf{S}(\Gamma)\). Since \(\Gamma\) is a scalar variable, this subproblem is solved numerically using a bounded one-dimensional search. According to \(\bar{\gamma}_e(\eta,\Gamma)=\frac{1+\eta}{2^{\Gamma}}-1\ge0\), the feasible interval of \(\Gamma\) is given by \(0\le\Gamma\le\log_2(1+\eta)\). The one-dimensional search is further restricted to the positive-definite barrier domain, i.e.,
$\mathcal{I}_\Gamma=\left\{\Gamma\in[0,\log_2(1+\eta)]:\mathbf{S}(\Gamma)\succ0\right\}$,
and therefore \(\Gamma\) is updated as
\begin{equation}
	\Gamma^{(t+1)}
	=
	\arg\max_{\Gamma\in\mathcal{I}_\Gamma}
	\left\{
	\Gamma+\mu\log\det\mathbf{S}(\Gamma)
	\right\}.
	\label{eq:update_gamma}
\end{equation}

\paragraph{Update $\boldsymbol{\phi}_c$}
For fixed $\lambda$, $\Gamma$, and $\eta$, the RIS phase-shift
subproblem maximizes
$F_\mu(\boldsymbol{\phi}_c)=\mu\log\det
\mathbf S(\boldsymbol{\phi}_c)$ over the complex circle product
manifold
$\mathcal M=\{\boldsymbol{\phi}_c\in\mathbb C^{N_R}:
|\phi_{c,n}|=1,\ \forall n\}$.
The Wirtinger gradient $\nabla_{\boldsymbol{\phi}_c^*}F_\mu$
is derived in Appendix~\ref{app:gradient}. Under the real inner
product $\langle\mathbf u,\mathbf v\rangle
=\Re(\mathbf u^H\mathbf v)$, the Euclidean and Riemannian
gradients are given by
$\mathbf g_E=2\nabla_{\boldsymbol{\phi}_c^*}F_\mu$,
$\operatorname{grad}F_\mu=\operatorname{Proj}_{\boldsymbol{\phi}_c}(\mathbf g_E)$,
where
$\operatorname{Proj}_{\boldsymbol{\phi}}(\mathbf u)
=\mathbf u-\Re(\overline{\boldsymbol{\phi}}\odot\mathbf u)
\odot\boldsymbol{\phi}$.
The conjugate search direction is updated as
$\mathbf d^{(t)}
=\operatorname{grad}F_\mu(\boldsymbol{\phi}_c^{(t)})
+\beta_t\mathcal T_{\boldsymbol{\phi}_c^{(t-1)}
	\rightarrow\boldsymbol{\phi}_c^{(t)}}(\mathbf d^{(t-1)})$,
where $\beta_t$ is the conjugate coefficient and $\mathcal T$
denotes vector transport. If $\mathbf d^{(t)}$ is not an ascent
direction, it is reset to $\operatorname{grad}F_\mu$.
The phase vector is then updated by retraction:
$
\boldsymbol{\phi}_c^{(t+1)}
=R_{\boldsymbol{\phi}_c^{(t)}}(\alpha_t\mathbf d^{(t)})
$,
where $\alpha_t$ is determined by Armijo backtracking to ensure
a non-decreasing barrier objective while maintaining
$\frac{1}{\sigma_u^2}
\left|\mathbf h_u^H(\boldsymbol{\phi}_c^{(t+1)})
\mathbf w_c\right|^2\geq\eta$,
$\mathbf S(\boldsymbol{\phi}_c^{(t+1)},\lambda,\Gamma,\eta)
\succ0$.
Here, $R_{\boldsymbol{\phi}_c^{(t)}}(\cdot)$ denotes the
retraction mapping onto $\mathcal M$.

\paragraph{Update \(\eta\)}
For fixed \((\boldsymbol{\phi}_c,\Gamma,\lambda)\), within the positive-definite barrier domain, the barrier objective is non-decreasing in \(\eta\), since increasing \(\eta\) enlarges the admissible Eve-SINR threshold \(\bar{\gamma}_e(\eta,\Gamma)\) and thus enlarges the bottom-right entry of \(\mathbf{S}\). Hence, the largest feasible value of \(\eta\) is optimal, which is given in closed form by
$\eta = \frac{1}{\sigma_u^2}|\mathbf{h}_u^{H}(\boldsymbol{\phi}_c)\mathbf{w}_c|^2\label{eq:update_eta}$.

The above block-coordinate updates are repeated within each inner iteration, while the barrier parameter \(\mu\) is gradually reduced in the outer loop.

\subsection{Convergence and Computational Complexity Analysis}

The convergence of the proposed algorithm is analyzed separately for the two stages. In Stage~I, the closed-form MRT solution globally maximizes the effective illumination power for fixed RIS phase shifts, while the RCG method with an Armijo line search ensures non-decreasing objective updates for fixed beamforming. Hence, $f_s^{(t+1)}\geq f_s^{(t)}$. Since the effective illumination power is upper bounded by the transmit-power constraint, the sequence $\{f_s^{(t)}\}$ converges.
In Stage~II, for fixed beamforming variables and barrier parameters, the block updates of $\lambda$, $\Gamma$, $\boldsymbol{\phi}_c$, and $\eta$ ensure a non-decreasing logarithmic barrier objective. If this objective is bounded above, the inner-loop objective sequence converges. To ensure monotonicity of the overall AO procedure, an incumbent feasible solution is maintained and updated only when a feasible candidate achieves a worst-case secrecy rate no lower than that of the incumbent. Consequently, $R_c^{(t+1)}\geq R_c^{(t)}$. Since the worst-case secrecy rate is upper bounded under the transmit-power constraint, the incumbent objective sequence converges. Nevertheless, the SDR relaxation, Gaussian randomization, and non-convex RIS phase-shift optimization generally lead to a suboptimal solution.

The computational complexity mainly arises from the sensing optimization in Stage~I and the robust secure transmission optimization in Stage~II. In Stage~I, the beamforming subproblem admits a closed-form MRT solution, while the RIS phase shifts are optimized using RCG. The resulting complexity is approximately
$\mathcal{O}\!\left(
	I_{\mathrm{AO}}^{(s)}
	\left[
	M_tN_R+I_{\mathrm{RCG}}^{(s)}N_R
	\right]
	\right)$,
where $I_{\mathrm{AO}}^{(s)}$ and $I_{\mathrm{RCG}}^{(s)}$ denote the numbers of outer AO and inner RCG iterations, respectively.
In Stage~II, the beamforming subproblem is solved via SDP feasibility tests with bisection, while the RIS phase shifts are optimized using an RCG-based logarithmic barrier method. The resulting complexity is approximately
$\mathcal{O}\!\left(I_{\mathrm{AO}}^{(c)}
\left[I_{\mathrm{bis}}I_{\mathrm{IP}}\mathcal{C}_{\mathrm{SDP}}(M_t,d)+I_{\mathrm{RCG}}^{(c)}(N_Rd^2+d^3)\right]\right)$,
where $I_{\mathrm{bis}}$, $I_{\mathrm{IP}}$, and $I_{\mathrm{RCG}}^{(c)}$ denote the numbers of bisection, interior-point, and RCG iterations, respectively, and $d=4$ is the dimension of the S-procedure-induced LMI. With fixed $d$, the RCG-based phase-shift optimization scales linearly with $N_R$, whereas the SDP-based beamforming subproblem generally dominates the computational cost for large $M_t$.

\section{Simulation Results}
Unless otherwise specified, the simulation parameters are set as follows. The carrier frequency and system bandwidth are set to
$f_c=30$~GHz and $B=20$~MHz, respectively, and the antenna spacing is
set as $d=\lambda/2$. The BS is equipped with a $4\times4$ UPA, while the hybrid RIS consists of a $4\times8$ array, where the reflection region contains $N_R=4\times4$ adjustable reflecting elements.
The locations of the BS, RIS, Bob, and Eve are respectively set as
$\mathbf p_{BS}=(0,0,5)$, $\mathbf p_{RIS}=(10,10,5)$,
$\mathbf p_b=(15,5,0)$, and $\mathbf p_e=(17,4,3)$.
The Rician factors of the BS-RIS, BS-Bob, and RIS-Bob links are set to
$\mathcal{K}=10$, while those of the BS-Eve and RIS-Eve links are set to
$\mathcal{K}=5$. The path-loss exponent is set to $\beta=2.2$. The BS transmit power is set to $P=20$~W, and the noise power is set to
$-70$~dBm. The number of sensing snapshots scales approximately linearly
with the sensing time fraction $\tau$, according to
$L=\lfloor T_{\mathrm{frame}}f_s\tau\rfloor$, yielding
$L=1024$ at $\tau=0.15$. The confidence level is set to $\alpha=0.95$.
The convergence tolerance of both optimization stages is set to $10^{-4}$.

\begin{figure}[htbp]
	\centering
	\includegraphics[width=0.4\textwidth]{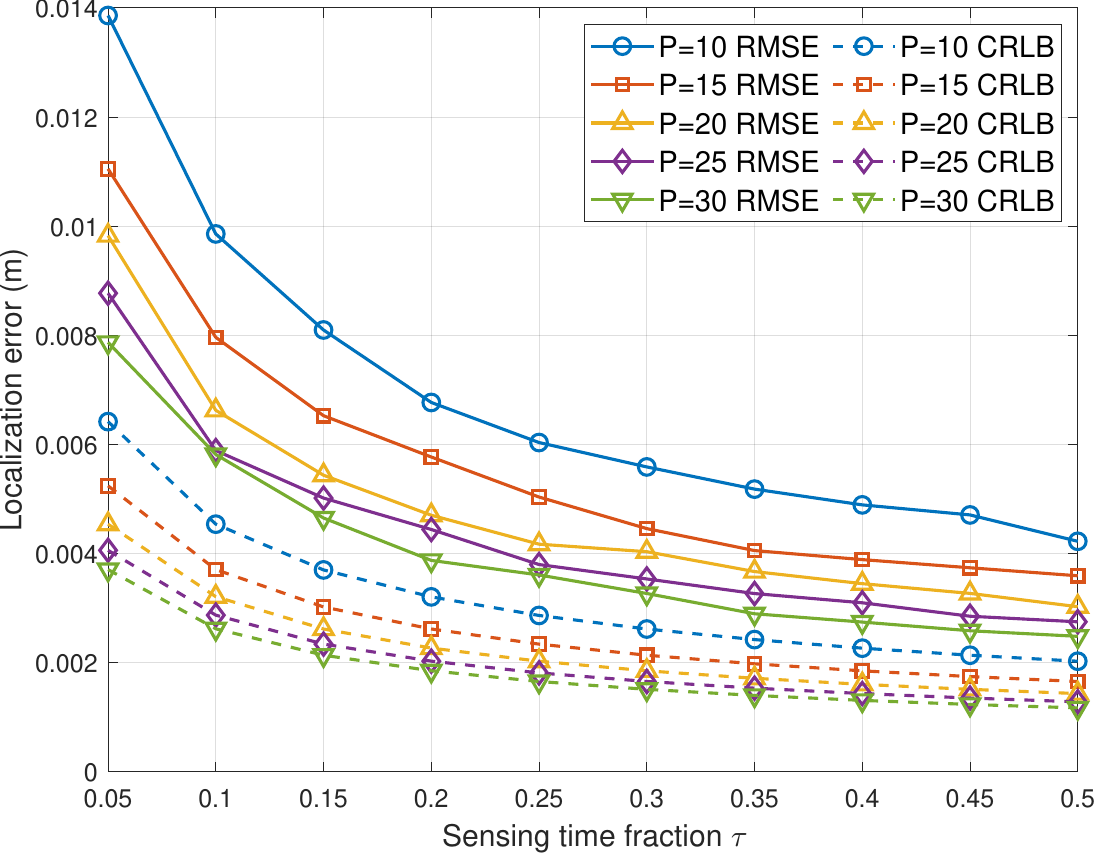}
	\caption{CRLB and RMSE of Eve localization versus sensing time fraction under different transmit powers.}
	\label{fig2}
\end{figure}

To mitigate the impact of random channel realizations on the simulation results, 5000 Monte Carlo trials are conducted under different transmit power levels and sensing time fractions $\tau$. As shown in Fig.~\ref{fig2}, increasing $\tau$ provides more sensing snapshots in Stage I, thereby reducing the nominal CRLB and the localization RMSE. However, the marginal performance improvement gradually diminishes as more sensing resources are allocated. Based on the Monte Carlo localization results, $\kappa=4.5$ is adopted as an empirical scaling factor, yielding an observed uncertainty-region coverage probability of approximately $95\%$ under the considered simulation settings. This value is fixed in the subsequent simulations.

\begin{figure}[htbp]
	\centering
	\includegraphics[width=0.4\textwidth]{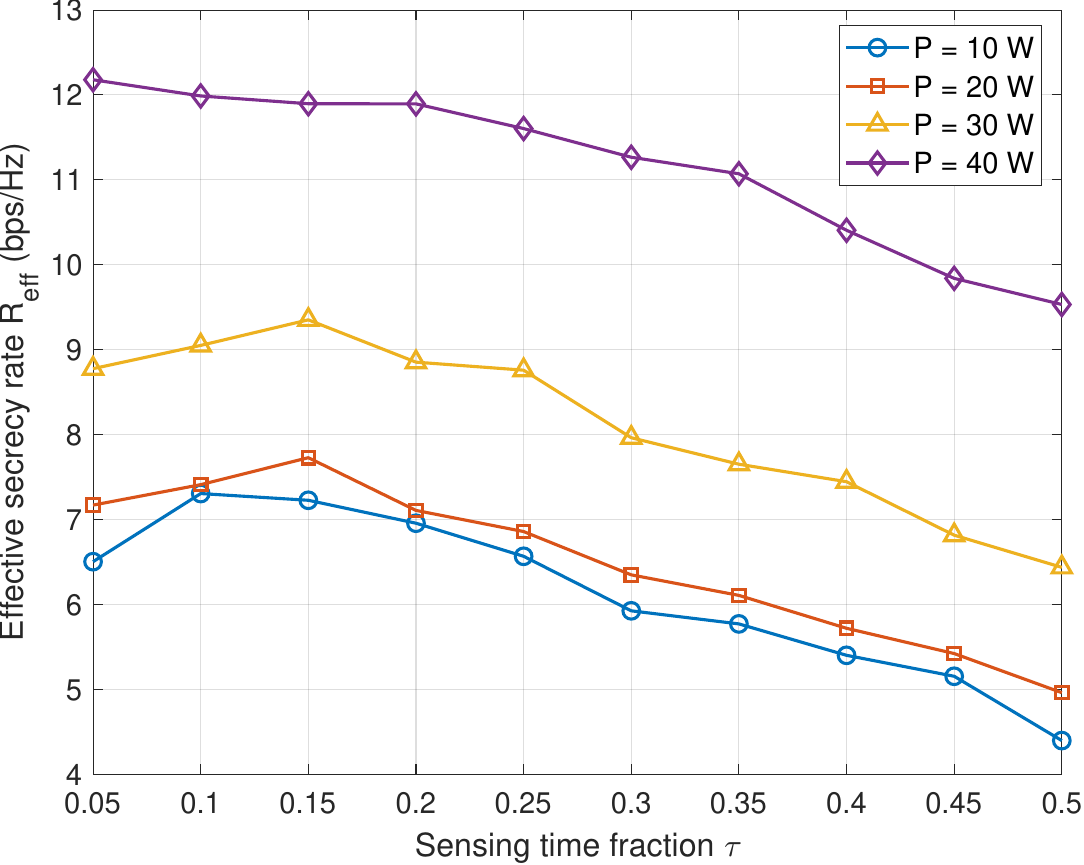}
	\caption{Secrecy rate versus $\tau$ under different parameters.}
	\label{fig3}
\end{figure}

As shown in Fig.~\ref{fig3}, the sensing time fraction $\tau$ introduces a tradeoff between localization accuracy and effective secrecy rate. For $P\leq30$~W, $R_{\rm eff}$ first increases and then decreases with $\tau$, reaching its peak at approximately $\tau^\star=0.10$--$0.15$. Increasing the sensing duration improves Eve's localization accuracy, thereby reducing the location uncertainty region and enhancing the worst-case secrecy performance. However, an excessively large $\tau$ reduces the available transmission time, causing the transmission-time loss to dominate. For $P=40$~W, $R_{\rm eff}$ generally decreases with $\tau$, indicating that the additional secrecy gain from sensing is insufficient to compensate for the loss in transmission time. Therefore, $\tau=0.15$ is adopted in the remaining simulations.

\begin{figure}[htbp]
	\centering
	\includegraphics[width=0.44\textwidth]{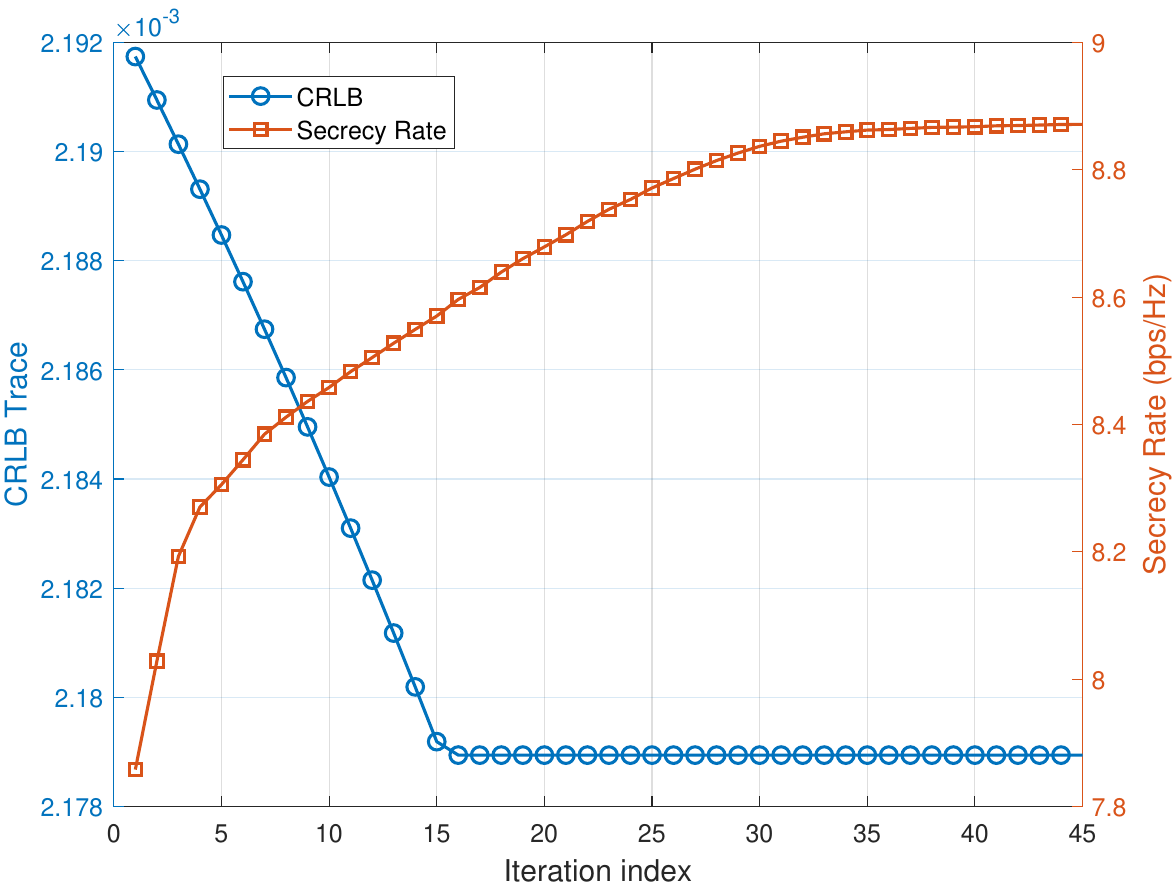}
	\caption{Convergence performance of the proposed alternating optimization algorithm.}
	\label{fig4}
\end{figure}

After determining the sensing time fraction $\tau$ and the uncertainty inflation factor $\kappa$, the convergence performance of the proposed AO algorithm is shown in Fig.~\ref{fig4}. As the number of iterations increases, the CRLB decreases while the worst-case secrecy rate increases, and both eventually converge. The sensing optimization converges within a relatively small number of iterations, whereas the secrecy-rate optimization requires more iterations due to the coupling among the beamforming vector, RIS phase shifts, and Eve's location uncertainty. These results demonstrate that the proposed AO algorithm achieves stable convergence within a limited number of iterations.

\begin{figure}[htbp]
	\centering
	\includegraphics[width=0.4\textwidth]{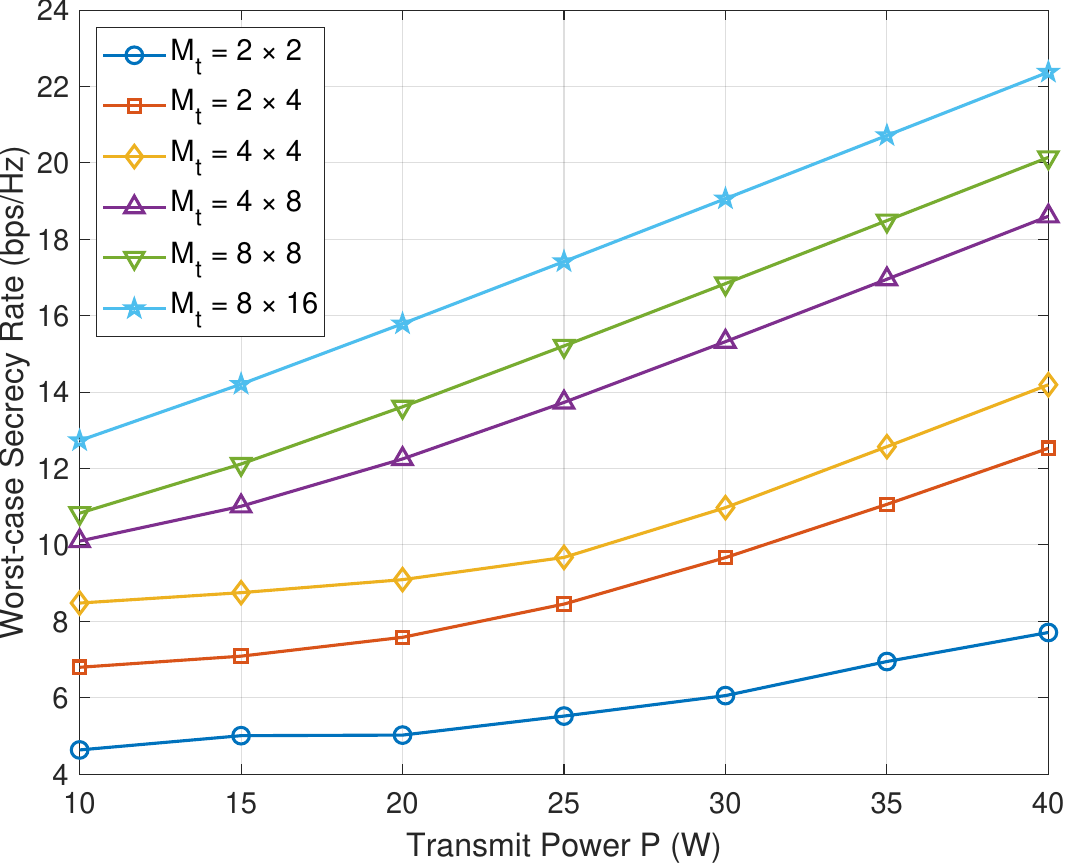}
	\caption{Impact of BS Transmit Power and Antenna Configuration on the Worst-Case Secrecy Rate.}
	\label{fig5}
\end{figure}

As shown in Fig.~\ref{fig5}, the worst-case secrecy rate increases with the BS transmit power for all antenna configurations, since higher transmit power improves the received signal quality at Bob. Moreover, increasing the BS antenna size leads to higher secrecy rates. For a small-scale array, e.g., a $2\times2$ UPA, the performance gain is limited by the insufficient spatial degrees of freedom for simultaneously enhancing Bob's signal and suppressing information leakage toward Eve under location uncertainty. With more antennas, the increased array gain and spatial degrees of freedom enable more effective beamforming optimization, resulting in higher worst-case secrecy rates. These results highlight the importance of the BS spatial degrees of freedom in robust RIS-assisted ISAC secure transmission.

\begin{figure}[htbp]
	\centering
	\includegraphics[width=0.4\textwidth]{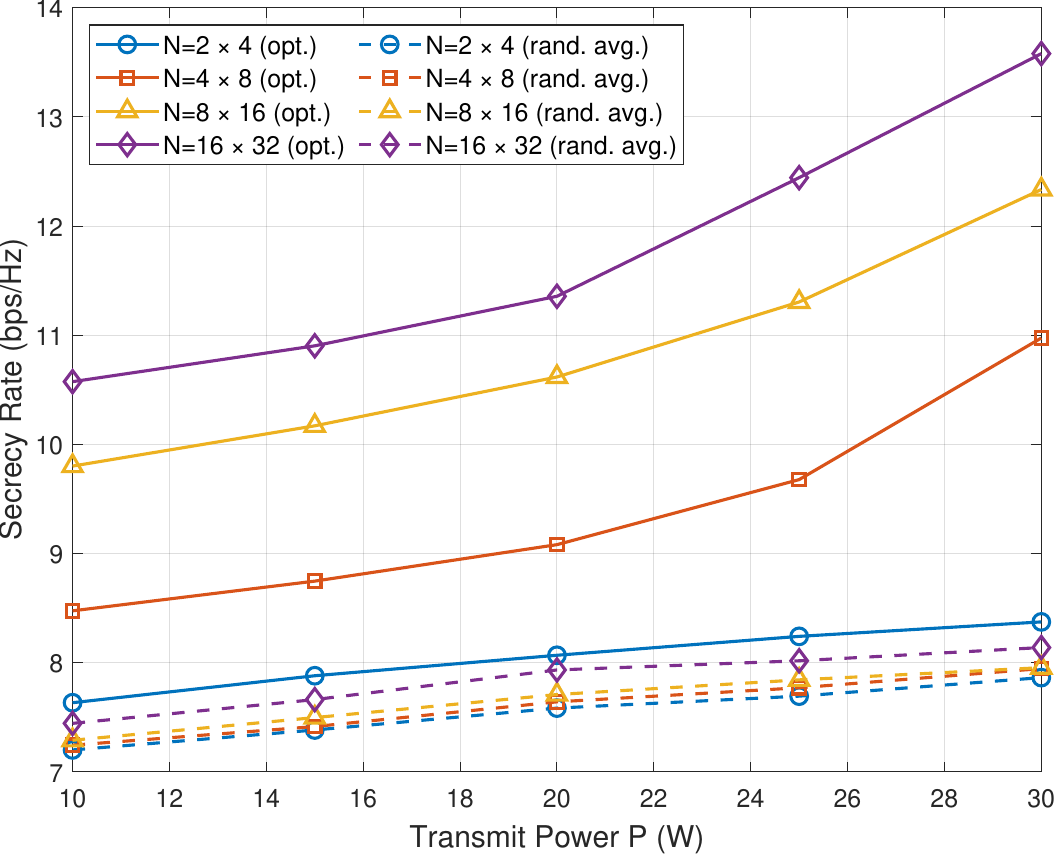}
	\caption{Worst-case secrecy rate versus RIS size under different transmit powers.}
	\label{fig6}
\end{figure}

As shown in Fig.~\ref{fig6}, the worst-case secrecy rate increases with the BS transmit power for all RIS configurations. Moreover, increasing the number of RIS elements further improves the secrecy performance by providing more reflection and phase-control degrees of freedom for enhancing Bob's signal and suppressing information leakage toward Eve. Compared with random RIS phase shifts, the proposed optimized RIS phase design achieves higher secrecy rates across different RIS sizes and transmit power levels, demonstrating the effectiveness of jointly optimizing the RIS phase shifts and transmit beamforming for robust secure transmission.

\begin{figure}[htbp]
	\centering
	\includegraphics[width=0.4\textwidth]{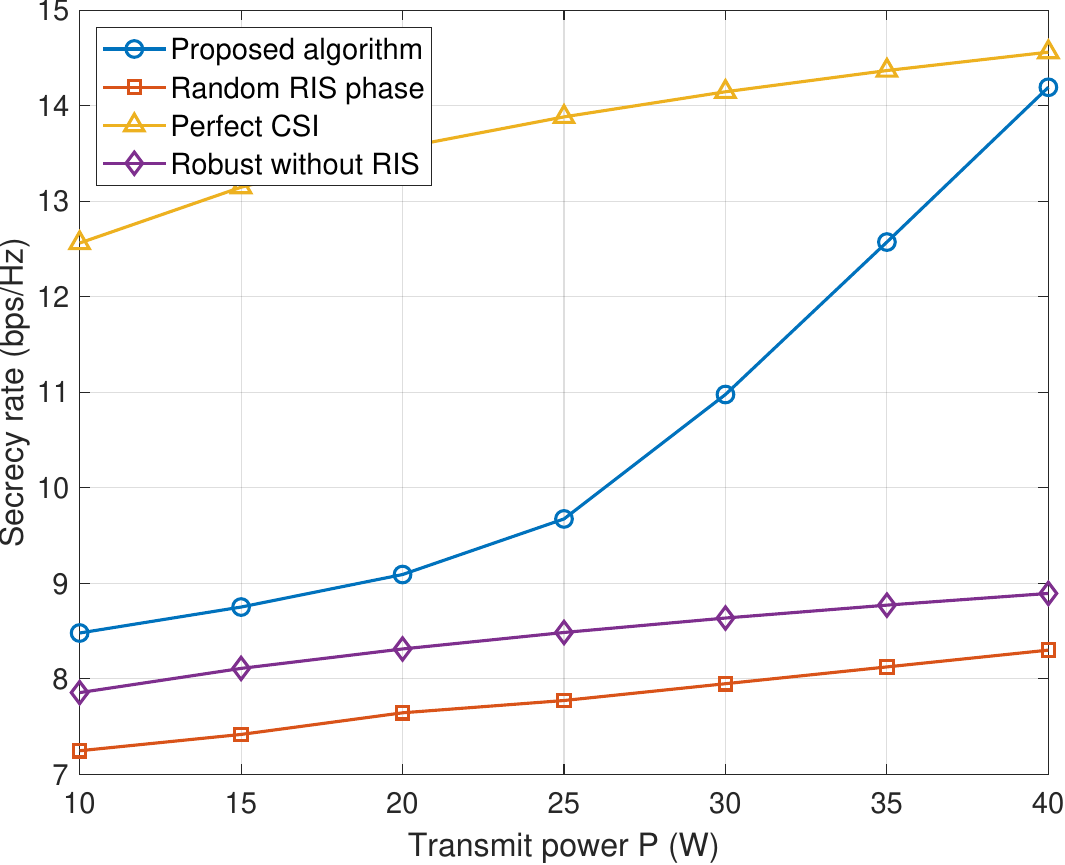}
	\caption {Secrecy Rate Versus Transmit Power Under Different Schemes.}
	\label{fig7}
\end{figure}

Fig.~\ref{fig7} shows the secrecy rate versus transmit power for different schemes. For all schemes, the secrecy rate increases with transmit power, and the perfect-CSI benchmark achieves the highest secrecy rate since it is not affected by Eve's location uncertainty. In the low-power regime, the advantage of the proposed scheme over the no-RIS and random-RIS schemes is not yet pronounced. As the transmit power increases, the proposed scheme improves more rapidly and gradually outperforms both, indicating that it can better exploit the spatial degrees of freedom at higher transmit power. In the high-power regime, the proposed scheme approaches the perfect-CSI benchmark, demonstrating that the proposed robust optimization can effectively mitigate the performance loss caused by Eve's location uncertainty.

\section{Conclusion}
This paper has investigated robust secure transmission in RIS-assisted ISAC systems with Eve location uncertainty and proposed a sensing-assisted two-stage optimization framework. In the first stage, the sensing beamformer and RIS reflection coefficients have been jointly optimized for accurate Eve localization, based on which an ellipsoidal location uncertainty region has been constructed using the localization CRLB. In the second stage, the communication beamformer and RIS reflection coefficients have been jointly optimized to maximize the worst-case secrecy rate, where the robust constraints have been reformulated into tractable LMI forms. Simulation results have demonstrated the effectiveness of the proposed approach and revealed a tradeoff between the sensing time fraction and the effective secrecy rate. Moreover, increasing the BS antennas, transmit power, and RIS elements can improve the worst-case secrecy rate, while the proposed method outperforms the benchmark schemes. Future work will extend the proposed framework to more practical and complex scenarios, including multiple legitimate users, multiple non-colluding eavesdroppers, and multiple RISs.

\appendices
\section{Gradient Derivation for RIS Phase Optimization in Stage I}
\label{app:gradient_sensing}
Since $g(\boldsymbol{\phi}_s)$ is a real-valued function with respect to
the complex variable $\boldsymbol{\phi}_s$, the Euclidean gradient is
obtained according to Wirtinger calculus as
\begin{equation}
	\nabla g(\boldsymbol{\phi}_s)
	=
	-2
	(c_0+\boldsymbol{\phi}_s^T\mathbf v)
	\mathbf v^{*}.
	\label{eq:app_euclidean_gradient}
\end{equation}
The feasible set
$
\mathcal M
=
\{
\boldsymbol{\phi}_s\in\mathbb C^{N_R}
\mid
|\phi_{s,n}|=1,\forall n
\}
$
forms a complex circle product manifold.
The tangent space at
$\boldsymbol{\phi}_s$ is given by
\begin{equation}
	T_{\boldsymbol{\phi}_s}\mathcal M
	=
	\{
	\mathbf d\in\mathbb C^{N_R}
	\mid
	\Re(\overline{\phi}_{s,n}d_n)=0,\forall n
	\}.
\end{equation}
The Euclidean gradient is projected onto the tangent space to obtain the
Riemannian gradient
\begin{equation}
	\operatorname{grad}
	g(\boldsymbol{\phi}_s)
	=
	\operatorname{Proj}_{\boldsymbol{\phi}_s}
	(
	\nabla g(\boldsymbol{\phi}_s)
	),
\end{equation}
where the projection operator is defined as
\begin{equation}
	\operatorname{Proj}_{\boldsymbol{\phi}_s}
	(\mathbf u)
	=
	\mathbf u
	-
	\Re
	(
	\overline{\boldsymbol{\phi}}_s
	\odot
	\mathbf u
	)
	\odot
	\boldsymbol{\phi}_s .
	\label{eq:app_projection}
\end{equation}
To update the manifold point, the retraction operation is adopted
\begin{equation}
	R_{\boldsymbol{\phi}_s}(\mathbf d)
	=
	\frac{
		\boldsymbol{\phi}_s+\mathbf d
	}
	{
		|
		\boldsymbol{\phi}_s+\mathbf d
		|
	},
	\label{eq:app_retraction}
\end{equation}
where the division is performed element-wise.
The vector transport between two tangent spaces is implemented by
\begin{equation}
	\mathcal T_{\boldsymbol{\phi}_s\rightarrow\boldsymbol{\psi}_s}
	(\mathbf d)
	=
	\operatorname{Proj}_{\boldsymbol{\psi}_s}(\mathbf d).
	\label{eq:app_transport}
\end{equation}
Based on the Riemannian gradient, the conjugate search direction is
updated using the Hestenes-Stiefel (HS) formula. Specifically,
$\beta_j=\frac{\left\langle\mathbf g^{(j)}-\mathbf g_{\rm old},\mathbf g^{(j)}\right\rangle}{\left\langle\mathbf g^{(j)}-\mathbf g_{\rm old},
\mathbf d_{\rm old}\right\rangle}$,
where the inner product is defined as
$
	\langle\mathbf a,\mathbf b\rangle
	=
	\Re(\mathbf a^H\mathbf b)
$.
To improve numerical stability, the HS+ modification is adopted: $\beta_j=\max(\beta_j,0)$.
The search direction is then updated as
\begin{equation}
	\mathbf d^{(j)}
	=
	-\mathbf g^{(j)}
	+
	\beta_j\mathbf d_{\rm old}.
\end{equation}
The step size $\alpha_j$ is determined by the Armijo backtracking line
search. Starting from an initial value, it is reduced by a factor
$\tau\in(0,1)$ until
\begin{equation}
	g
	\left(
	R_{\boldsymbol{\phi}_s^{(j)}}
	(\alpha_j\mathbf d^{(j)})
	\right)
	\leq
	g(\boldsymbol{\phi}_s^{(j)})
	+
	c\alpha_j
	\langle
	\mathbf g^{(j)},
	\mathbf d^{(j)}
	\rangle ,
\end{equation}
where $c\in(0,1)$ is the Armijo parameter.

\section{Gradient Derivation for RIS Phase Optimization in Stage II}
\label{app:gradient}

For fixed $\mathbf W$, the equivalent Eve channel vector can be
written as
\[
\mathbf h_e(\mathbf p_e;\boldsymbol{\Phi}_c)
=
\mathbf h_{be}(\mathbf p_e)
+
(\mathbf H_{br}^{\mathcal R})^H
\boldsymbol{\Phi}_c^H
\mathbf h_{re}^{\mathcal R}(\mathbf p_e).
\]
Define
$
\mathbf D_n
\triangleq
\frac{\partial\mathbf h_0}{\partial\phi_{c,n}^{*}}
$, $
\mathbf G_n
\triangleq
\frac{\partial\mathbf J_h}{\partial\phi_{c,n}^{*}}
$.
Then, the derivatives of the nominal Eve channel and its
Jacobian with respect to $\phi_{c,n}^{*}$ are given by
$
	\mathbf D_n
	=
	(\mathbf H_{br}^{\mathcal R})^H
	\mathbf E_{nn}\mathbf h_{re}^{\mathcal R}$,
$	\mathbf G_n
	=
	(\mathbf H_{br}^{\mathcal R})^H
	\mathbf E_{nn}
	\frac{\partial\mathbf h_{re}^{\mathcal R}}
	{\partial\mathbf p_e^T}
$,
where $\mathbf E_{nn}$ denotes the elementary matrix whose
only nonzero entry is the $(n,n)$-th entry.

Let
$\widetilde{\mathbf W}=\mathbf W/\sigma_e^2$.
Since $\mathbf W$ is fixed in the RIS phase-shift subproblem,
the derivatives of the quadratic-form coefficients
$\mathbf A=\Re(\mathbf J_h^H\widetilde{\mathbf W}\mathbf J_h)$,
$\mathbf b=\Re(\mathbf J_h^H\widetilde{\mathbf W}\mathbf h_0)$,
and $c=\mathbf h_0^H\widetilde{\mathbf W}\mathbf h_0$
are obtained using Wirtinger calculus as
\begin{equation}
	\frac{\partial\mathbf A}{\partial\phi_{c,n}^{*}}
	=
	\frac{1}{2}
	\left(
	\mathbf J_h^H\widetilde{\mathbf W}\mathbf G_n
	+
	\mathbf G_n^T\widetilde{\mathbf W}^{*}\mathbf J_h^{*}
	\right).
\end{equation}
\begin{equation}
	\frac{\partial\mathbf b}{\partial\phi_{c,n}^{*}}
	=
	\frac{1}{2}
	\left(
	\mathbf J_h^H\widetilde{\mathbf W}\mathbf D_n
	+
	\mathbf G_n^T\widetilde{\mathbf W}^{*}\mathbf h_0^{*}
	\right).
\end{equation}
\begin{equation}
	\frac{\partial c}{\partial\phi_{c,n}^{*}}
	=
	\mathbf h_0^H\widetilde{\mathbf W}\mathbf D_n.
\end{equation}

Since $\bar{\gamma}_e(\eta,\Gamma)$ and $\mathbf M$ are
independent of $\boldsymbol{\phi}_c$, the derivative of the
LMI matrix with respect to $\phi_{c,n}^{*}$ is
\begin{equation}
	\frac{\partial\mathbf S}{\partial\phi_{c,n}^{*}}
	=
	\begin{bmatrix}
		-\dfrac{\partial\mathbf A}{\partial\phi_{c,n}^{*}}
		&
		-\dfrac{\partial\mathbf b}{\partial\phi_{c,n}^{*}}
		\\[6pt]
		-\left(\dfrac{\partial\mathbf b}
		{\partial\phi_{c,n}^{*}}\right)^T
		&
		-\dfrac{\partial c}{\partial\phi_{c,n}^{*}}
	\end{bmatrix}.
\end{equation}
For $\mathbf S\succ0$, the Wirtinger derivative of the
logarithmic barrier function
$F_\mu(\boldsymbol{\phi}_c)=\mu\log\det\mathbf S$
is given by
\begin{equation}
	\frac{\partial F_\mu}{\partial\phi_{c,n}^{*}}
	=
	\mu\operatorname{tr}
	\left(
	\mathbf S^{-1}
	\frac{\partial\mathbf S}{\partial\phi_{c,n}^{*}}
	\right).
\end{equation}
Collecting the derivatives of all RIS elements yields the
Euclidean gradient
\begin{equation}
	\nabla_{\boldsymbol{\phi}_c^*} F_\mu = \mu \left[ \operatorname{tr}\left( \mathbf S^{-1} \frac{\partial\mathbf S}{\partial\phi_{c,1}^{*}} \right), \ldots, \operatorname{tr}\left( \mathbf S^{-1} \frac{\partial\mathbf S}{\partial\phi_{c,N_R}^{*}} \right) \right]^T.
\end{equation}
Under the real inner product
$\langle\mathbf u,\mathbf v\rangle=\Re(\mathbf u^H\mathbf v)$,
the Euclidean gradient is
$\mathbf g_E=2\nabla_{\boldsymbol{\phi}_c^{*}}F_\mu$.
\balance
\bibliographystyle{IEEEtran}  % 选择参考文献格式
\bibliography{references}     % 引用 .bib
\end{document}